\documentclass[aip,jcp,reprint]{revtex4-1}
\usepackage{bm}
\usepackage{amsmath}
\usepackage{graphicx}
\usepackage{amssymb,chemarr}
\usepackage[usenames]{color}
\usepackage{braket,calc}
\usepackage{amsfonts}
\usepackage{mathrsfs} % for a script C in an equation
\usepackage{xspace}
\begin{document}

% Hagan commands\
\newcommand{\ie}{\textit{i.e.}\xspace}
\newcommand{\eg}{\textit{e.g.}\xspace}

\newcommand{\sub}[1]{ _{\mathrm{#1}}}
\newcommand{\kbt}{k_\mathrm{B} T}
\newcommand{\kb}{k_\mathrm{B}}
\newcommand{\kt}{k_\mathrm{B}T}
\newcommand{\bx}{\mathbf{x}}
\newcommand{\str}{\mathcal{S}}
% Perkett commands
\newcommand{\cp}{\mathrm{CP}_2}
\newcommand{\Ncv}{N_\mathrm{cv}}
\newcommand{\ABS}{ CC$\rightleftharpoons$AB\xspace}
\newcommand{\ABTR}{ CC:TR$\rightleftharpoons$AB:TR\xspace}

% editing commands
\definecolor{Blue}{rgb}{0,0.0,1.0}
\definecolor{Red}{rgb}{0.7,0.0,0.0}
\definecolor{Green}{rgb}{0,1.0,0}
\definecolor{Purple}{rgb}{0.27,0,0.51}

\newcommand{\mrp}[1]{\textcolor{Blue}{#1}}
\newcommand{\revision}[1]{\textcolor{Red}{#1}}
\newcommand{\mfh}[1]{\textcolor{Green}{#1}}
\newcommand{\dtm}[1]{\textcolor{Blue}{#1}}
\newcommand{\comment}[1]{\textcolor{Green}{#1}}
\newcommand{\changec}[2]{\textcolor{Red}{#1}\textcolor{Blue}{#2}}
\newcommand{\schange}[2]{{\sout{#1}}{\change{#2}}}

\title{The Allosteric Switching Mechanism in Bacteriophage MS2}
\author{Matthew R. Perkett}
\thanks{These authors contributed equally}
\author{Dina T. Mirijanian}
\thanks{These authors contributed equally}
%\affiliation{Martin Fisher School of Physics, Brandeis University, Waltham, MA, USA}
\author{Michael F. Hagan}
\email{hagan@brandeis.edu}
\affiliation{Martin Fisher School of Physics, Brandeis University, Waltham, MA, USA}

\begin{abstract}
In this article we use all-atom simulations to elucidate the mechanisms underlying conformational switching and allostery within the coat protein of the bacteriophage MS2. Assembly of most icosahedral virus capsids requires that the capsid protein adopt different conformations at precise locations within the capsid.  It has been shown that a 19 nucleotide stem loop (TR) from the MS2 genome acts as an allosteric effector, guiding conformational switching of the coat protein during capsid assembly. Since the principal conformational changes occur far from the TR binding site, it is important to understand the molecular mechanism underlying this allosteric communication. To this end, we use all-atom simulations with explicit water combined with a path sampling technique to sample the MS2 coat protein conformational transition, in the presence and absence of TR-binding.  The calculations find that TR binding strongly alters the transition free energy profile, leading to a switch in the favored conformation. We discuss changes in molecular interactions responsible for this shift.  We then identify networks of amino acids with correlated motions to reveal the mechanism by which effects of TR binding span the protein.  The analysis predicts amino acids whose substitution by mutagenesis could alter populations of the conformational substates or their transition rates.
\end{abstract}

\maketitle

%%% Introduction %%%
\section{Introduction}

The controlled interconversion between protein conformational states is crucial for essential cellular functions, including signaling, metabolism, and assembly of the dynamic cytoskeleton. A key regulatory role in such processes is often played by allosteric effectors, whose binding favors a particular protein conformation.  The transition pathways by which proteins interconvert between these folded states are largely unknown because intermediates along the pathways cannot be directly characterized by experiments. Similarly, it remains poorly understood how perturbations due to effector binding are communicated across the protein to alter its conformational free energy landscape. In this article we combine long unbiased all-atom molecular dynamics (MD) simulations, an efficient pathway sampling algorithm called the string method \cite{Weinan2002,E2007,E2010,Weinan2005,Maragliano2006,Ovchinnikov2011,Pan2008,Ovchinnikov2014,Maragliano2014}, and analysis of inter-residue correlations to characterize a protein conformational transition pathway and how it is affected by effector binding.  In particular, we study the conformational transition of the MS2 coat protein dimer, and how the binding of an RNA stem loop from the MS2 genome acts as a molecular conformational switch that guides protein assembly into an icosahedrally symmetric capsid.

\begin{figure*}[htbp]
  \begin{center}
    \includegraphics[width=\textwidth]{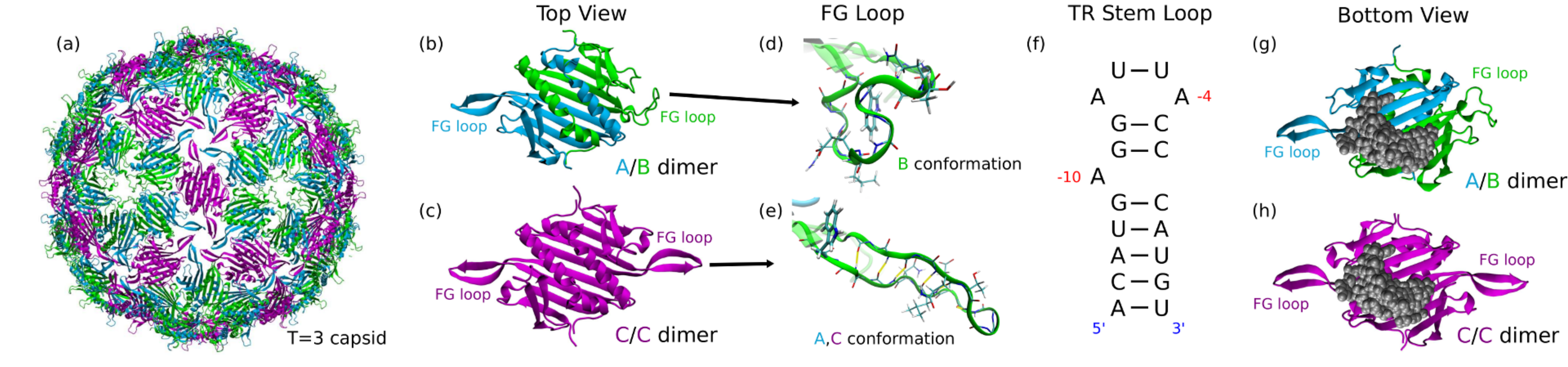}
  \caption{MS2 Capsid geometry and subunit structure. \textbf{(a)} The complete T=3 MS2 capsid of 27.5nm diameter is comprised of 30 CC and 60 AB dimers.  It has icosahedral symmetry with the 5-fold vertices as AB dimers and the 3-fold vertices as 3 AB + 3 CC dimers. (pdb ID: 1BMS)  \textbf{(b)-(c)} the AB and CC dimers colored according to their conformation.  The B conformation differs significantly from the A and C conformations in the FG loop.  \textbf{(d)-(e)} A close up view of the FG loop with a selection of side chains shown as bonds.  The B conformation lacks the hydrogren bonds found in the A and C conformations (and shown in yellow).  \textbf{(f)} The nucleic acid sequence of the TR stem loop, which binds with high affinity to the base of the MS2 dimer.  The sequence positions of the adenines that bind most strongly are labeled in red (-10 and -4).  \textbf{(g)-(h)} MS2 AB and CC dimers shown with the RNA stem loop (TR) bound to their base (pdb ID: 2BU1 \cite{Grahn2001}).  The RNA can adopt two symmetric positions for the CC dimer (only one shown), but the AB dimer allows only one position due to steric collisions. The RNA is shown as grey VDW spheres.}
  \label{fig:ms2}
  \end{center}
\end{figure*}

MS2 is a small bacteriophage that infects male \textit{E. Coli}. During virus assembly, 180 copies of the coat protein (CP) spontaneously assemble around a 3,569 nucleotide single-stranded RNA genome to form an icosahedral capsid. The capsid is a $T{=}3$ structure, meaning that the CPs adopt three conformations (termed A,B,C) which are precisely arranged within the capsid \cite{Caspar1962}. Major structural differences among the protein conformations are confined to the FG loop, which in the A and C conformations forms an anti-parallel $\beta$-hairpin, but in the B conformation is a flexible loop pulled back against the dimer with a small $\alpha$-helix kink.  The A and C monomers are thus nearly identical, and their FG loops meet at 20 3-fold (quasi-6-fold) axes, whereas the FG loops of the B monomers meet at the 12 5-fold interfaces. In solution, the monomers form stable, non-covalent dimers, which are the basic assembly subunits and will be denoted as $\cp$ (Fig. \ref{fig:ms2} b,c).  Formation of the capsid thus requires that 30 CC and 60 AB dimers associate and arrange themselves into the icosahedral geometry (Fig. \ref{fig:ms2} a).

% FIGURE

Based on structural studies, in vitro assembly assays, and modeling, it has been proposed that allosteric interactions between $\cp$ and the viral genome guide conformational selection during MS2 assembly \cite{Stockley2007,Dykeman2010,ElSawy2010,Morton2010}.  Capsid assembly can be triggered \textit{in vitro} by the addition of a 19-nucleotide RNA stem-loop (TR) fragment from the genome.  TR encompasses the start codon for the replicase protein, and has been shown to bind strongly to the bottom of $\cp$ \cite{Gott1993,Stockley1995}.  In the crystal structure, TR is bound to the CC dimers in two symmetric orientations, while steric constraints allow only a single orientation for the AB dimer (Fig. \ref{fig:ms2} g,h).

\textit{In vitro} experiments by Stockley and coworkers  \cite{Stockley2007} on wild-type CP showed that, in the absence of genomic RNA, CP assembles slowly and produces only a low yield of capsids.  Adding a molar ratio of TR results in a strongly bonded $\cp$:TR complex that is kinetically trapped.  However, adding an equal molar ratio of $\cp$ to $\cp$:TR results in rapid and efficient assembly. Furthermore, NMR studies on an assembly-incompetent mutant MS2 coat protein (Trp82Arg), showed that TR binding induces a conformation change from a symmetric dimer (presumably BB-like) to an asymmetric dimer (presumably AB-like).  

Based on these observations, it was proposed that during assembly of wild-type (WT) MS2 capsid proteins, TR binding acts as a molecular switch which favors a conformational change from the symmetric CC dimer to the asymmetric AB dimer \cite{Stockley2007}.
Since both AB and CC dimers are needed for efficient assembly, this scenario is consistent with the observation that pure solutions of either $\cp$ (assumed to be CC) or $\cp$:TR (assumed to be AB:TR) are kinetically trapped whereas an equal molar ratio of $\cp$ to $\cp$:TR results in rapid and efficient assembly. Subsequent theoretical models suggest that such a conformational switch is consistent with existing structural data and assembly kinetics~\cite{Dykeman2010,ElSawy2010,Morton2010,Dykeman2011}.

Since TR binds $\cp$ \cite{Stockley1995} (Fig.~\ref{fig:ms2} g,h) about 12 \AA\ from the FG loop where the conformation change is localized, there is great interest in understanding the molecular mechanism underlying the apparent allosteric communication between these two regions of the protein. Using all-atom normal mode analysis, Dykeman et al.\cite{Dykeman2010} found that TR binding to an initially symmetric CC conformation leads to asymmetries consistent with the AB conformation. Namely, fluctuations of residues near the FG loop on the A$^*$ chain (meaning the chain that corresponds to the A chain in the AB dimer conformation) are suppressed, whereas those near the B$^*$ FG loop increase.

The goal of this paper is to directly calculate the MS2 capsid protein conformational free energy landscape, to learn how it is altered by the binding of the genome fragment TR, and to elucidate the molecular basis by which perturbations caused by TR binding are communicated across the protein. To this end, we employ the string method \cite{Weinan2002,E2007,E2010,Weinan2005,Maragliano2006} to identify and characterize the most probable transition pathways and associated free energy profiles for the conformational transition in the presence and absence of TR, using all-atom simulations with explicit water. Furthermore, to directly probe the molecular basis for allosteric communication, we characterize correlations of amino acid conformational statistics and motions within long, unbiased MD simulation trajectories. These combined calculations demonstrate that the conformational transition is a complex, multi-step process with multiple metastable minima, and is stabilized by multiple molecular-scale interactions whose statistics can be altered by molecular binding in disparate regions of the protein. The analysis predicts several amino acids whose substitution by mutagenesis could alter populations of the conformational substates or their transition rates. These findings may shed light on the mechanisms by which molecular binding affects conformational free energy landscapes in a wide variety of proteins, as well as for understanding the diverse roles of RNA in viral assembly.

Previous computational works have used enhanced sampling methods to examine the effect of small molecule substrates on protein interconversion pathways and free energies, with a particular focus on the enzyme adenylate kinase \cite{Seyler2014,Arora2007,Lu2014,Matsunaga2012,Wang2013}. Most closely related to our work, Pattis and May investigated the effect of RNA binding on the Lassa Virus nucleoprotein conformational free energy landscape\cite{EricMay1}. 

This article is arranged as follows. In section~\ref{sec:methods} we describe the model, simulations, and methodologies used to sample the transition. In section~\ref{sec:results}, we describe the transition pathways predicted by the string method in the presence and absence of TR, we highlight some residues found to play key roles in stabilizing the transition based on the converged strings, and we present results of mutual information on correlations between amino acid conformations and motions. Finally, in section~\ref{sec:discussion} we discuss implications of these results for understanding the mechanism underlying the conformational transition and how it is influenced by TR binding.  Additional methodological details and validations are given in the appendices.

\section{Methods}
\label{sec:methods}

\subsection{Systems and simulations}
\label{subsec:system}
{\bf Systems.} For statistical analysis and for generating beginning and end points for string method calculations, we initialized unbiased MD simulations from two MS2 capsid protein dimer conformations, each in the presence and absence of the RNA stem loop TR. We denote the four systems as AB, CC, AB:TR, and CC:TR.  To avoid complications associated with the fact that P78 undergoes a \emph{cis} to \emph{trans} switch between conformations, we studied P78N mutants, which assemble complete capsids but are not infectious~\cite{Hill1997}.  The AB and CC dimer structures were therefore extracted from a crystal structure of the empty P78N capsid (pdb ID 1BMS \cite{Stonehouse1996}).  Since no crystal structure for P78N capsids with TR is available, we extracted AB:TR and CC:TR from a wild type MS2 capsid containing TR (pdb ID 2BU1), and performed the P78N mutation in silico using VMD \cite{Humphrey1996}.  The first and last bases in the RNA stem loop (A and U) for 2BU1 are missing, and were added using CHARMM~\cite{Brooks2009}.

Each of the four dimer structures was solvated with at least 1nm of water on each side of the structure. The resulting simulation boxes were approximately 10.2nm x 7.7nm x 5.6nm for $\cp$ and 10.6nm x 7.2nm x 7.5nm for $\cp$:TR. We ensured that each pair of systems intended to serve as beginning and end points of the same string (AB, CC) and (AB:TR, CC:TR) had the same number of atoms.  Water molecules were replaced at random with Na$^+$ and Cl$^-$ ions to neutralize the charge and to bring the ionic strength to 0.1M.  The total system size was approximately 41,000 atoms for $\cp$ and 58,000 atoms for $\cp$:TR. During equilibration, an orientation restraint was added to keep the dimer from self-interaction across the periodic boundary. For long unbiased MD calculations, larger water boxes (approximately 10.2nm$^3$ for both $\cp$ and $\cp$:TR) were used with no orientation restraints. Details about the equilibration protocol are given in Appendix~\ref{appendix:subsec:ms2_equilibration}.

{\bf Simulations.} Simulations were performed with version 4.5.5 of Gromacs~\cite{VanDerSpoel2005} modified with version 1.3.0 of the plugin PLUMED~\cite{Bonomi2009}, which was used to generate all restraints and monitor collective variables. Version 5.0.5 of Gromacs was used for the long unbiased simulations.  The CHARMM36 all-atom forcefield~\cite{Best2012} was used to represent the system and the TIP3P model~\cite{Jorgensen1983,Price2004} was used for water molecules (The string simulations used the CHARMM22/CMAP forcefield \cite{MacKerell1998,Mackerell2004} for proteins as they were partly performed before the CHARMM36 forcefield was available in Gromacs). Bond lengths were constrained using the LINCS algorithm~\cite{Hess1997} with order 4.  The NPT ensemble was simulated using velocity rescaling for the temperature coupling and the Parrinello-Rahman barostat  for pressure coupling \cite{Parrinello1981,Nose1983}.  Electrostatic interactions were calculated using the particle-mesh Ewald (PME) algorithm~\citep{Essmann1995}, with a grid spacing of 0.12 and real-space interactions cut off at 1.2nm.  Van der Waals interactions were switched at $1.0$nm and cut off at $1.2$nm.

\subsection{The String Method Algorithm}
\label{subsec:string_method}

To determine the minimum free energy transition pathways (MFTP)  for the AB$\rightleftharpoons$CC and AB:TR$\rightleftharpoons$CC:TR conformations, we used the string method algorithm in collective variables, which was first presented by Maragliano et al\cite{Maragliano2006}. While a number of powerful methods have been developed to sample transition pathways and other rare events (e.g., \cite{Rohrdanz2013,Dickson2010,Bolhuis2002,Pan2008,Ovchinnikov2011,Fischer2011,Elber2007,Pietrucci2009,Lei2004,Moroni2004,Allen2005,Pfaendtner2009,Barducci2010,Zhang2007,Huber1996,Ferguson2011,VanErp2012,Bowman2011,Noe2009,ChenTomo,Elber2016,Maragliano2015}), the string method provides a means to discover the MFTP in a space of many collective variables (CVs), with a computational expenditure that is nearly independent of the number of CVs.  To obtain a meaningful free energy minimum, the collective variables must include all slow degrees of freedom relevant to the transition. Our collective variables were chosen to be a subset of the atomic positions \cite{Ovchinnikov2011,Lacroix2012,Zhao2013,Ovchinnikov2014}.

The method can be summarized as follows.  A set of collective variables capable of characterizing all relevant slow degrees of freedom in the system is chosen (section~\ref{subsubsec:selecting_CVs}).  An initial pathway connecting the two stable states is discretized as an ordered sequence of states (called images) and represented as a curve (called a string) in the multidimensional collective variable space. 
An iterative calculation is then performed to relax the initial pathway toward a minimum free energy pathway: (1) For each image, multiple short MD simulations are performed in which sampling is constrained to the vicinity of that image in collective variable space by a harmonic bias potential.  The gradient of the free energy in collective variable space at each image is calculated from the average force imposed by the bias potential. (2) The position of each image in collective variable space is incremented by displacing it along the (negative) direction of the free energy gradient. (3) Images are redistributed to maintain uniform spacing in arc length along the string. Steps (1-3) are repeated until the string converges to within desired precision. The free energy profile along the converged string can then be calculated using umbrella sampling (section~\ref{subsec:string_free_energy}).

 %We outline our implementation here;
 Details about these procedures, the chosen set of coordinates, and assessments of convergence are given in Appendix~\ref{appendix:string}.%Further details can be found in in appendix~\ref{appendix:string} and Refs~\cite{Weinan2002,E2007,E2010,Weinan2005,Maragliano2006,Ovchinnikov2011,Pan2008,Ovchinnikov2014,Maragliano2014}.

\subsection{Selecting Collective Variables}
\label{subsubsec:selecting_CVs}

A string is defined by a set of collective variables (CVs), which must include all slow degrees of freedom that are relevant to the reaction. It is not known  \emph{a priori} which CVs constitute a good reaction coordinate. While in principle it is possible to choose a large number of CVs with the expectation that a subset of them will constitute a good reaction coordinate, extraneous or redundant CVs can introduce noise that slows convergence.  Thus, our goal was to select the minimal possible CV set sufficient to describe both the $\cp$ and $\cp$:TR conformational transitions.

Based on extensive trial calculations using various types of CVs (including distances, positions, and dihedral angles), we found Cartesian positions of individual atoms to be best suited for the study of MS2.  Atomic positions have been successfully used in previous studies \cite{Ovchinnikov2011,Lacroix2012,Zhao2013,Ovchinnikov2014} and can capture both the native CC backbone hydrogen bonds breaking/forming and the formation of the $\alpha$-kink in the AB state.

Because absolute positions are not invariant under rigid body motions, we restricted translational and rotational diffusion by including position restraints on 10 C$_\alpha$ atoms in the top helices of each monomer in $\cp$ (residues 105-109).  These residues are far from the RNA binding site and FG loop (where the conformational change is localized). In an alternate approach, Ovchinnikov et al. performed principle component analysis on the rigid core of the protein to define a body-centered coordinate system~\cite{Ovchinnikov2011}.  Another approach is to perform on the fly structural alignment~\cite{Branduardi2013}.  

\begin{figure}[hbtp]
  \begin{center}
    \includegraphics[width=\columnwidth]{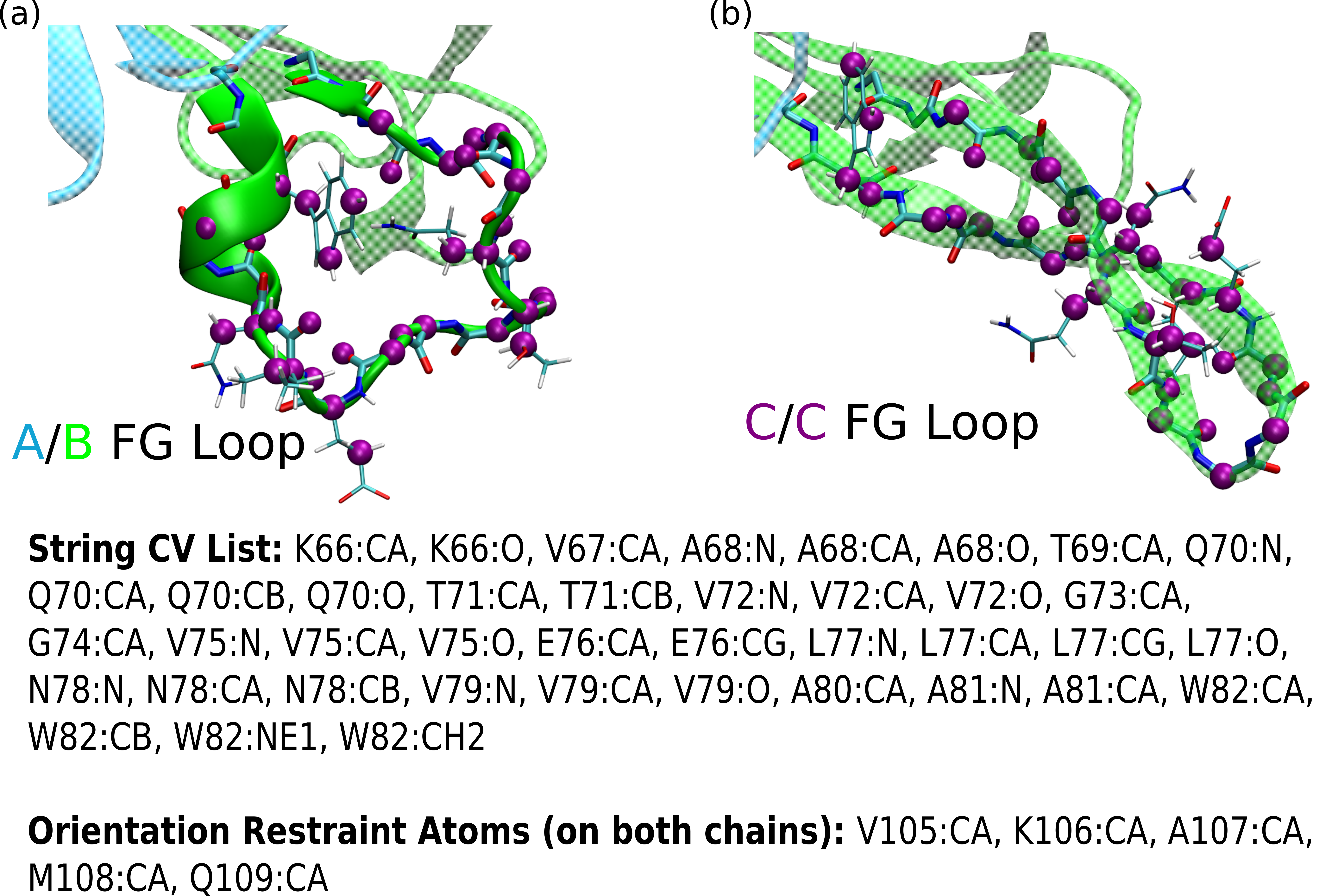}
  \caption{The atoms whose positions were selected as CVs for the string are shown for both the AB and CC FG loops.  The backbone and a selection of the side chains are shown as bonds, and the string atoms are shown as purple spheres.  The string atoms are listed in the text below the figure along with those atoms whose positions were used to prevent translational and rotational diffusion.}
  \label{fig:ms2_cvs}
  \end{center}
\end{figure}

We followed the approach of Ref.~\cite{Ovchinnikov2011} to select the set of atoms whose positions comprise the CVs. We ran a series of targeted molecular dynamics simulations (TMDs)~\cite{Schlitter1994}, in which external biasing forces were applied to the candidate atoms to force the system between conformations.  Each candidate set of atoms was ranked by the difference in backbone dihedral angles between the final structure and the target, and the amount of RMSD drift observed during 4ns of simulation after all restraints were released. This test was performed on TMD simulations in both directions (AB to CC and CC to AB) for both $\cp$ and $\cp$:TR.  Once a set of CVs was chosen in this manner, redundant or extraneous atoms were eliminated through a trial and error process in which candidate removals were tested by additional TMD simulations. The final set of CVs contains the positions for 40 atoms, listed in Fig~\ref{fig:ms2_cvs}.

% FIGURE

\subsection{Free Energy Along String Pathway}
\label{subsec:string_free_energy}
To calculate the free energy profiles from converged strings, we performed umbrella sampling on an order parameter $s$ that gives the position along the string path.  Our implementation is based on the approach described in Ref~\cite{Pontiggia2015}.  To ensure that sampling does not meander arbitrarily far in directions transverse to the transition tube defined by the string \cite{Ovchinnikov2014}, we also defined an order parameter $z$, which measures the distance from the string.  The definition of $s$ and $z$ are inspired by the path collective variables in PLUMED~\cite{Bonomi2009}; for an arc length between images $i$ and $i+1$, $s$ and $z$ are given by%
\begin{align}
s &= i + \frac {(\bm{y} - \boldsymbol{\theta}_i) \cdot (\boldsymbol{\theta}_{i+1} - \boldsymbol{\theta}_i)} {|\boldsymbol{\theta}_{i+1} - \boldsymbol{\theta}_i|^2} \nonumber \\
z^2 &= \frac {|\bm{y} - \boldsymbol{\theta}_i|^2} {|\boldsymbol{\theta}_{i+1}-\boldsymbol{\theta}_i|^2} - (s-i)^2
\label{eq_pathcv}
\end{align}
where $\boldsymbol{\theta}_i$ gives the vector of CV coordinates defined by image $i$, $\bm{y}$ is the dynamic vector of CV coordinates during sampling, and $s$ is the projection of $\bm{y}$ onto the line segment between the bounding images ($\boldsymbol{\theta}_{i+1} - \boldsymbol{\theta}_{i}$), scaled by the image separation.

During umbrella sampling approximately 150 window centers were spaced evenly in $s$, with a spring constant of $\kappa \approx 350$kJ. To maintain sampling near the center of the transition tube \cite{Ovchinnikov2014}, a half-harmonic, upper wall potential was placed between $z=2$ and $z=3$, with spring constant $\kappa_\text{wall} \approx 450$kJ.

To check for hysteresis, each window was seeded using steered MD simulations from two different starting structures, one from each of the upper and lower bounding images.  The two seeds for each of the $\approx$150 windows were then each sampled for 200ps, so the total simulation time for each free energy calculation was 60 ns. The free energy was calculated from this data using Alan Grossfield's implementation of WHAM (Weighted Histogram Analysis Method)~\cite{GrossfieldWham}.

\subsection{Root of Mean Squared Fluctuations (RMSF)}
\label{subsec:rmsf}
The root of mean squared fluctuations (RMSF) for each amino acid about an average structure were calculated for each of the $\cp$ and $\cp$:TR systems.  First, a single 300ns unbiased trajectory was run for each of the four systems and seed structures were extracted as starting points for new trajectories.  For the CC, CC:TR and AB:TR system each of the 7 seeded trajectories was 450ns long of which the last 368ns was used in the calculation, resulting in 2.576$\mu$s of sampling for each system.  For the AB system there were 6 seeded trajectories of 530ns length each, from which the last 430ns was used in the calculation, resulting in 2.58$\mu$s of sampling.  Configurations were outputted every 25ps for all the seeded unbiased MD trajectories.  The structures from the trajectories were first aligned to minimize the mass-weighted RMSD of the C$_\alpha$ atoms that comprise the core of the protein (residues 7-63 and 85-124 of each monomer).  Using the aligned structures, the RMSF was calculated with respect to the average structure and then averaged over all non-hydrogen atoms in each amino acid.

\subsection{Mutual Information}
\label{subsec:mut_info}
We calculated the mutual information (MI) between all pairs of amino acids for all $\cp$ and $\cp$:TR systems using the approach and  MutInf program developed by McClendon et al\cite{Mcclendon2010}.  In this approach, the MI is calculated using second order terms from the configurational entropy expansion, and indicates the correlation between backbone and side chain conformations\cite{Mcclendon2010}.  It is calculated using internal coordinates (i.e. the $\phi$ and $\psi$ backbone dihedrals and side chain rotamers).  Amino acids that have shared mutual information have correlated dihedral distributions.  Correlated distributions can arise through direct interaction, a chain of interactions, backbone movements, solvent rearrangement, or other mechanisms.

For each system, we applied the MutInf program to the microseconds of multiple trajectories we used in the RMSF calculation.  We used 24 bins per degree-of-freedom, and results from 20 sets of scrambled data were calculated and subtracted in order to determine the excess mutual information, as done by McClendon et al\cite{McClendon28102014}. We then used hierarchical clustering on the resulting MI matrix to identify groups of amino acids that share significant mutual information (as done in Ref~\cite{Mcclendon2010}).  We generated the dissimilarity matrix as given in Eq.~\ref{eq:dissim_mut_inf}, and used a Euclidean distance metric to cluster amino acids.

\begin{equation}
\label{eq:dissim_mut_inf}
D_{ij} = \mathrm{Max}(\mathrm{MI}) - \mathrm{MI}_{ij}
\end{equation}

We systematically extracted the largest possible ``real'' clusters by recursively splitting the hierarchy of clusters until each cluster achieved an MI average greater than a given cutoff value.  After generating the clusters, we verified that they were valid (i.e. had high intra-cluster MI averages and small  inter-cluster MI averages) using
\begin{equation}
\label{eq:clusterAvgs}
\mathrm{cluster\ avg} = \frac{1}{N} \sum_{i \in W_m} \sum_{j \in W_n, j \neq i} MI_{ij}
\end{equation}
where $W_m$ is the set of amino acid numbers that belong to cluster $m$.  The sums loop over all residues in $W_m$ and $W_n$, and $N$ is the total number of elements in the sum such that $i \neq j$.  For an intra-cluster average ($m=n$), all amino acid self-correlations are ignored (i.e. $i \neq j$).

From the MI network we calculated the node betweenness centrality for each residue.  The node betweenness centrality determines the number of shortest distance paths that go through each node of the network and is an indicator of the importance of that node in the communication of the network ~\cite{Boccaletti2006175}.  We used the \textit{tnet} package available through the statistical software R for the calculation of the betweenness centralities ~\cite{Opsahl2010}.

\section{Results}
\label{sec:results}

\subsection{Conformational Transition Pathway}
\label{subsec:string_results}
\begin{figure*}[btp]
  \begin{center}
    \includegraphics[width=\textwidth]{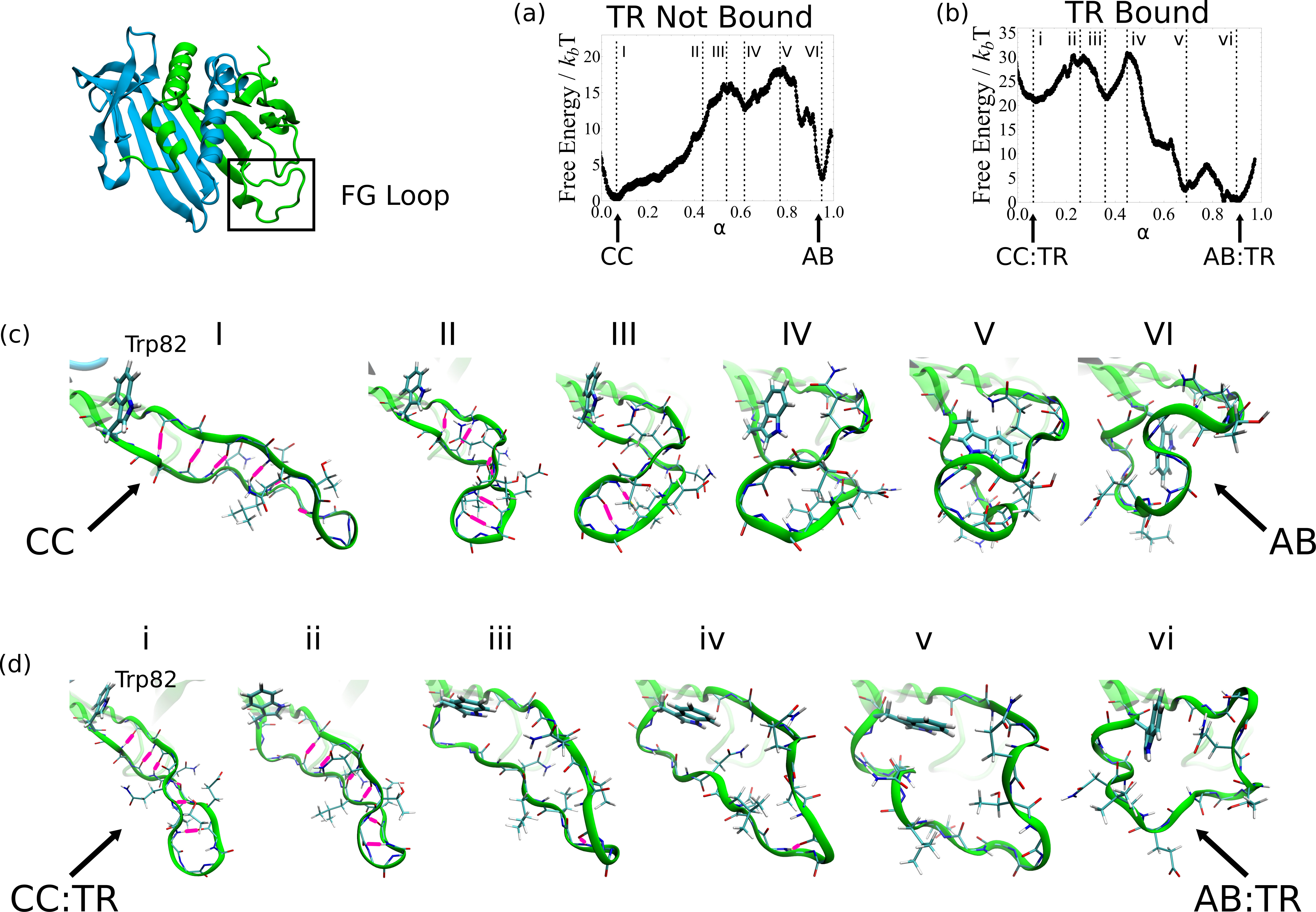}
   \caption{The most probable transition pathways and associated free energy profiles for \ABS and \ABTR.  \textbf{(a), (b)} The free energy along the most probable pathway as a function of arc length $\alpha$ along the converged strings.  \textbf{(c), (d)} Close up snapshots of the B$^*$ FG loop along the transition pathway for \ABS \textbf{(c)} and \ABTR \textbf{(d)}.  The native backbone hydrogen bonds of the CC monomer are shown in pink, and side chains with atoms selected as string collective variables (CVs) are shown as bonds.  The labels correspond to the position along the free energy profile as indicated in (a) and (b).}
  \label{fig:ms2_free_energy}
  \end{center}
\end{figure*}

In this section we compare the calculated most probable conformational transition pathways for the  \ABS and \ABTR  MS2 coat protein dimer interconversions. The free energy profiles calculated from the converged strings and illustrative snapshots from the converged strings are shown in Fig.~\ref{fig:ms2_free_energy}.
Details on these calculations can be found in Sections \ref{sec:methods}:A,B and Appendix \ref{appendix:string}. Furthermore, several tests of convergence are discussed in Appendix \ref{appendix:string_convergence}.  Most significantly, an independent string started from a different initial pathway produced a similar transition pathway and free energy profile (Fig.~\ref{fig:ms2_string_validation2}).

% FIGURE

\subsubsection{Pathway in the Absence of TR}
The \ABS calculation obtains that the symmetric CC state is favored over the AB state by a free energy of $\approx 3 \kt$, and there is one on-pathway metastable state.  The string pathway indicates the following order of events for a CC to AB transition with each number corresponding to the snapshots in Fig.~\ref{fig:ms2_free_energy}a,c:  The CC loop bends inward, straining the native backbone hydrogen bonds (I $\rightarrow$ II) and eventually breaking them (III), beginning with the bonds closest to the core of the protein (and Trp82).  After all of the native hydrogen bonds are broken, the FG loop opens becoming partially solvated and Trp82 leaves its hydrophobic pocket resulting in a metastable state (IV) with a free energy difference of about $12 \kt$ compared to the native CC structure.  The FG loop must now widen to accommodate further rotation of Trp82 (V) paying a $\approx 5 \kt$ free energy penalty before collapsing and rearranging into the final AB substate (VI).

\subsubsection{Pathway in the Presence of RNA}
Complexation of $\cp$ with the RNA stem loop TR dramatically shifts the free energy landscape, causing the AB:TR substate to be favored over the CC:TR substate by $\approx 20\kt$.  The transition pathway is also markedly different from \ABS, and now involves two on-pathway metastable states (Fig.~\ref{fig:ms2_free_energy}b,d).  In the converged string, the transition from CC:TR to AB:TR proceeds by the following sequence of events with each number corresponding to the snapshots in Fig.~\ref{fig:ms2_free_energy}b,d: The first two backbone hydrogen bonds near the base of the FG loop break (i $\rightarrow$ ii).  The backbone dihedral angles of amino acids 79-81 move toward their eventual position in the $\alpha$-kink of AB:TR.  This represents the first barrier to the transition, of $\approx 7\kt$.  It is now free energetically favorable for Trp82 to rotate out of the hydrophobic pocket toward chain A, which forms the first metastable state (iii).  Interestingly, this rotation proceeds in the opposite direction as found in the CC$\rightleftharpoons$AB transition.  A free energy penalty of $\approx 7\kt$ must be paid to reach state iv, which involves side chain rearrangements and further solvation of the FG loop.  Finally, Trp82 rotates into the FG loop, which then spontaneously collapses, resulting in the second metastable state (v). This state is structurally very similar to the final AB:TR state (vi), and only $\approx 2\kt$ higher in free energy.  A final rotation of Trp82, involving a $\approx 4\kt$ barrier, leads to the final AB:TR state (vi).

\subsubsection{Comparison of Pathways}
The most striking difference between the AB$\rightleftharpoons$CC and AB:TR$\rightleftharpoons$CC:TR strings is the shift in the most stable sub-state upon TR binding, from CC to AB:TR.  This population shift is consistent with experimental data~\cite{Stockley2007}.  Both pathways highlight the important role of the large side chain of Trp82 in determining the sequence of events during the conformation change. The strings show that the native CC backbone hydrogen bonds require much more substantial molecular rearrangements before breaking and allowing rotation of Trp82 in the \ABS transition as compared to \ABTR.  This difference suggests that the binding of TR destabilizes these hydrogen bonds, which would contribute to shifting the population toward the AB:TR state.

\subsection{Root of Mean Squared Fluctuations}
\label{subsec:msf_results}

\begin{figure*}[btp]
  \begin{center}
    \includegraphics[width=\textwidth]{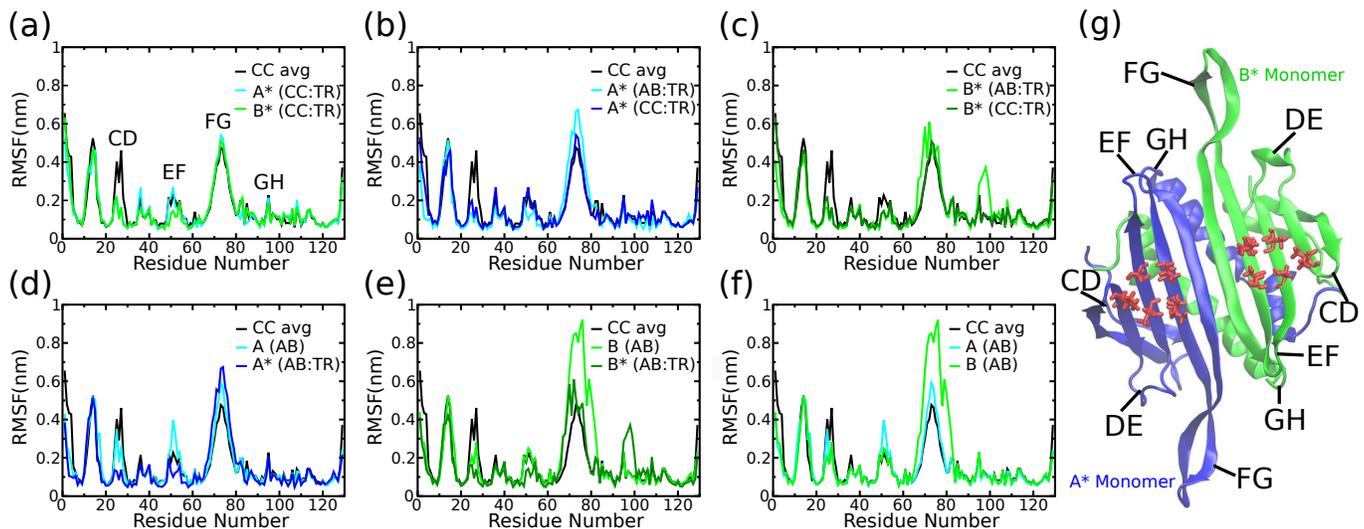}
   \caption{The RMSF for each residue (averaged over all non-hydrogen atoms within a residue) for each $\cp$ and $\cp$:TR system.  \textbf{(a)-(f)} The RMSF as a function of residue number with the important loops and turns labeled in \textbf{(a)}.  To facilitate comparison of the dynamics between different systems, each plot shows as a reference the RMSF for the CC dimer, averaged over the two symmetric monomers, as well as the two additional systems.  \textbf{(g)} A view of the CC dimer with the loops and turns labeled according to convention~\cite{Golmohammadi1996}.  The residues Val29, Thr45, Ser47, Thr59, Lys61 which form the TR binding pocket on each monomer are shown in red.}
  \label{fig:msf}
  \end{center}
\end{figure*}

The residue based RMSFs were calculated for the four dimer systems using microseconds of unbiased MD simulations for each system as described in Sec.~\ref{subsec:rmsf}.  The RMSF values give the typical fluctuations for each amino acid and provide insight into the relative flexibility of different portions of the protein.  These results are summarized in Fig.~\ref{fig:msf}.

Upon TR binding to the CC dimer, the largest change in the RMSF occurs in residues 23-30 of the CD loop, where the fluctuations decrease similarly in the A* and B* monomers  (Fig.~\ref{fig:msf}a).  Similarly, the loops in both the A* and B* monomers of the RNA bound systems have lower RMSF than those in the non-bound systems (Fig.~\ref{fig:msf}d,e).  The CD loop in the CC dimer has higher fluctuations than in any other system.  The effect of TR binding on the CD loop can be explained by noting that residues Asn27 and Val29 are in direct contact with TR.

While the CC dimer is symmetric, it is possible that the transition in the B* FG loop to the AB conformation is due in part to asymmetries that arise in dimer fluctuations upon TR binding.  The TR induced asymmetry in the fluctuations of the CC:TR system is evident in the RMSF of residues 49 to 53 leading up to and including part of the EF loop, which have higher RMSF in the A* monomer than in the B* monomer (Fig.~\ref{fig:msf}a).  This asymmetry is consistent with what was found in a previous all atom normal mode analysis by Dykeman et al. \cite{Dykeman2010}, which showed that for WT MS2 the B factor, which is directly related to the RMSF, of the EF loop decreased in B* and increased in A*.  The RNA binding to the AB dimer has the same effect on the EF loop of the B* monomer (Fig.~\ref{fig:msf}c).  This effect on the EF loop can be explained by the fact that this part of the protein in the B* monomer is in direct contact with the TR.  While Dykeman et al. also found that the B factor of the FG loop in B* increased upon TR binding to the CC dimer, we find the dynamics of the FG loops of the CC:TR system to be similar to the symmetric dimer.  However, the RMSF of residues Trp82, Arg83, Tyr85 and Leu86 following the FG loop are suppressed in the A* monomer (Fig.~\ref{fig:msf}a).

A significant difference between the CC:TR and AB:TR systems is in the FG loop, which has higher RMSF in the AB:TR system for both the A* and B* monomers (Fig.~\ref{fig:msf}b,c).  Another significant difference between the AB:TR and other systems is in the GH-loop of the B* monomers, where residues 95-100 have higher RMSF in B* monomer of AB:TR \ref{fig:msf}c,f.  The final effect to note is that TR binding in the AB system leads to a dramatic decrease in the dynamic fluctuations of the FG loop of the B* monomer, where in the AB:TR system the dynamics of the loop is much more confined (Fig.~\ref{fig:msf}f).  It is important to note that these differences in fluctuations in the B* monomer take place far from the TR binding residues and are a result of long range allosteric communications.

To validate that the sampling is sufficient for this analysis we present the RMSF of the EF and FG loops of the CC:TR dimer using half of the total trajectory frames in the Appendix~\ref{appendix:rmsf_results}, and show that the results are the same as those using all the frames.  In the next section we look at how the changes in the dynamics of TR binding residues are communicated to the rest of the protein, including the FG loop.

\subsection{Mutual Information}
\label{subsec:mutual_info}

{\bf Cluster analysis to determine groups of correlated residues.}
To characterize residue conformational correlations, we also calculated the mutual information between the dihedral angles of all pairs of amino acids from the long unbiased trajectories. We then clustered residues based on their pairwise mutual information to identify groups of residues that are strongly correlated. The intra-cluster averages used as cutoffs for each system were: $0.035\kt$ for CC, $0.04\kt$ for AB, 0.07$\kt$ for CC:TR and 0.1$\kt$ for AB:TR.  For further discussion on the cutoffs and the resulting number of clusters see Appendix~\ref{appendix:mutinfo_results}.  To assess sampling convergence, we compared the raw MI data and calculated clusters from all 7 trajectories  (2.576$\mu$s) to results obtained using only 4 trajectories (1.472$\mu$s).  The Pearson's correlation coefficient for the MI calculated from the full and partial data sets is 0.896, and clusters obtained from the partial data set (Appendix~\ref{appendix:mutinfo_results}) are consistent with those from the larger data set.

\begin{figure}[btp]
  \begin{center}
    \includegraphics[width=\columnwidth]{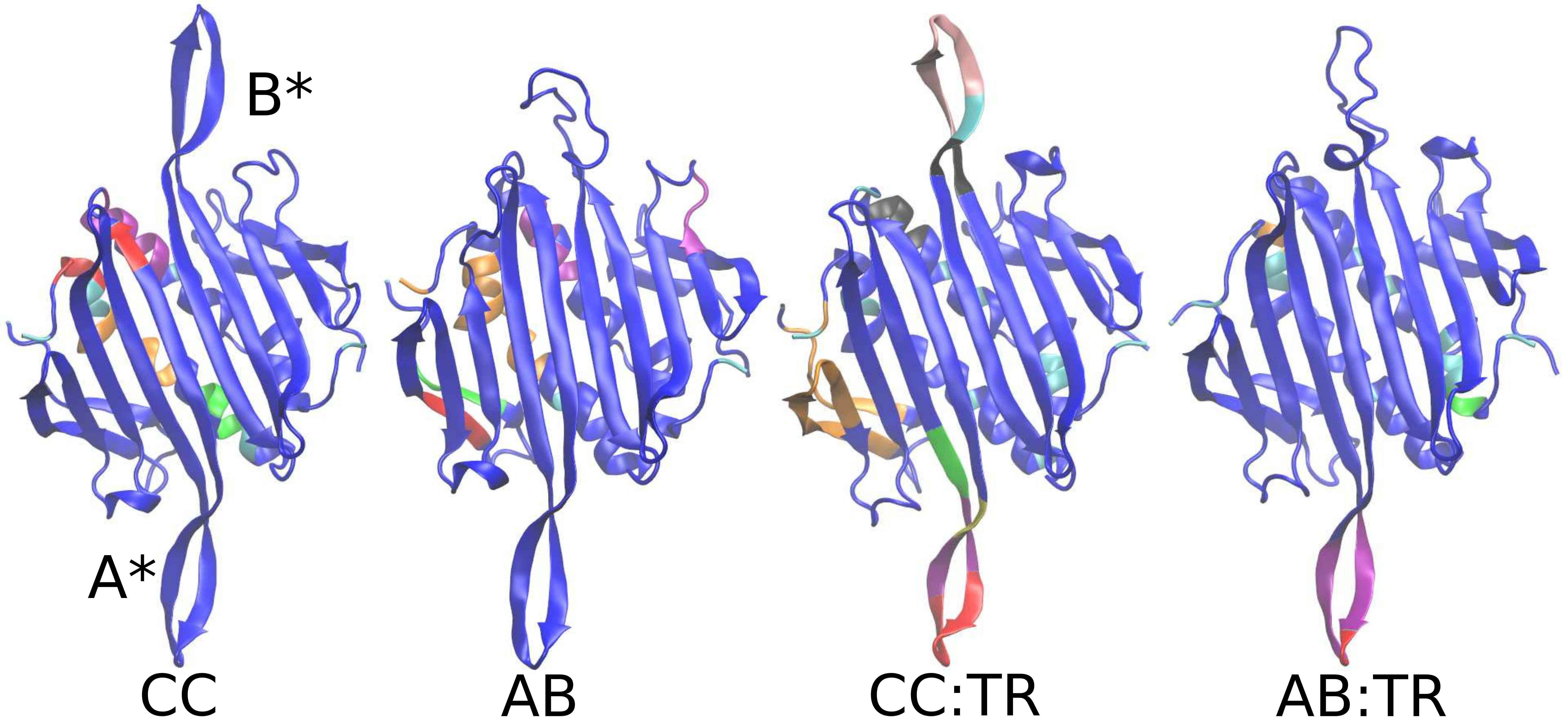}
   \caption{Mutual information clusters for all the $\cp$ and $\cp$:TR systems.   The clusters of correlated residues, calculated from the mutual information between pairs of amino acids as described in the text, are shown in different colors.}
  \label{fig:MI1}
  \end{center}
\end{figure}

The clusters from the full data set are presented in Fig.~\ref{fig:MI1} for each of the dimer systems. In the CC and AB dimers, there is one large cluster that encompasses the majority of residues on the TR binding face of the protein (blue in Fig.~\ref{fig:MI1}), including both the A* and B* FG loops.  Small clusters of residues form at the opposite face of the protein, each primarily containing few residues from the helical domains.

% FIGURE

\begin{figure}[hbtp]
  \begin{center}
    \includegraphics[width=\columnwidth]{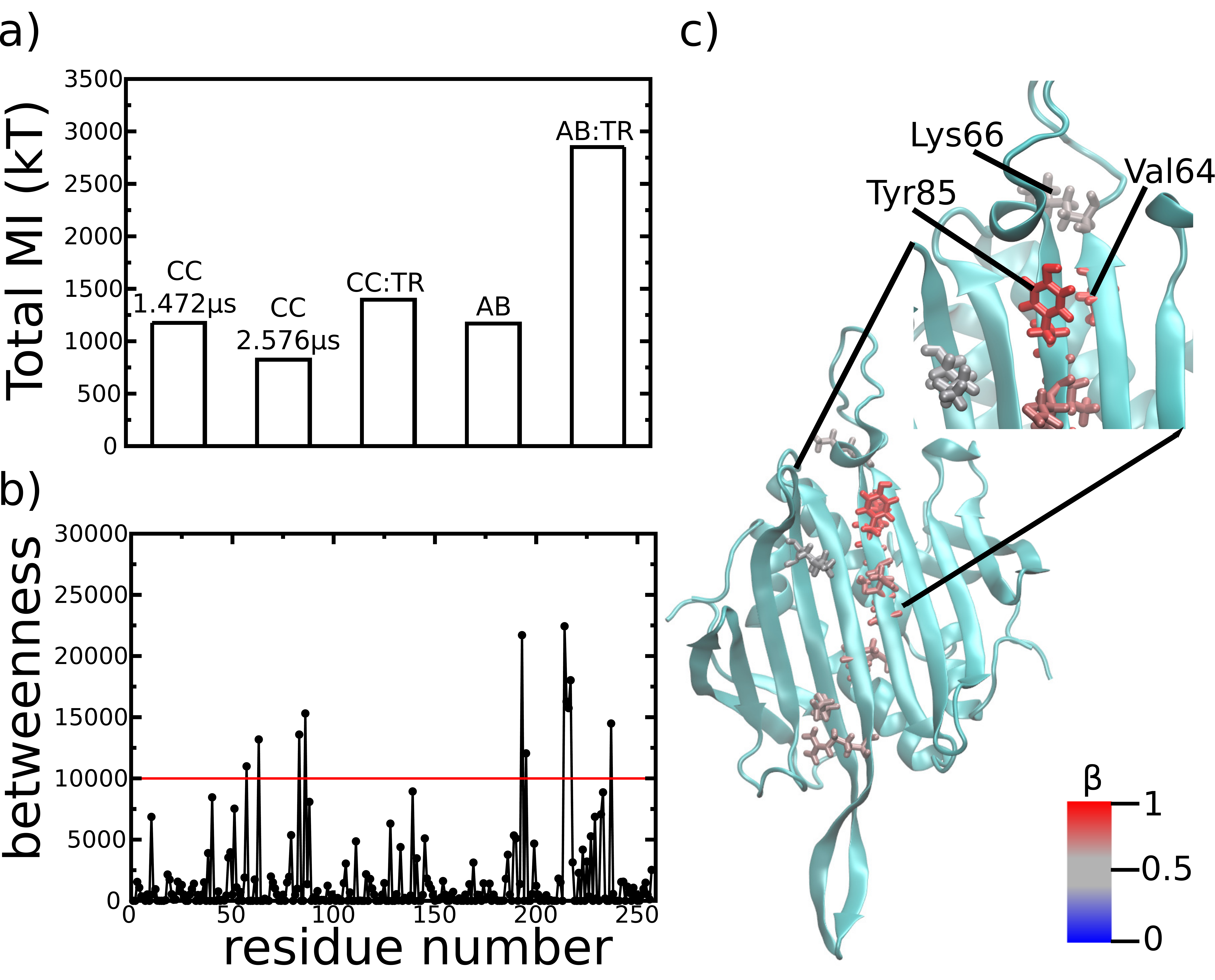}
   \caption{Total MI of each $\cp$ and $\cp$:TR system and the betweenness centralities of the AB:TR residues.  \textbf{(a)} The total MI of each system is compared.  Calculation of the total MI of the CC system for two different sampling sizes gives an estimate of the spread in total MI values due to sampling size variations. The AB:TR system is found to have stronger correlations than the other three systems.  \textbf{(b)} The betweenness centralities of each residue in the AB:TR system.  Residue numbering starts at the A* monomer and ends at the B* monomer.  \textbf{(c)} The residues with betweenness centralities greater than $10^4$ are represented on a structure of the AB:TR system in stick representation.  The color scale corresponds to the relative betweenness $\beta$, which is the ratio of the given residue's betweenness centrality divided by the maximum residue betweenness centrality in the system. The residues with betweenness centralities greater than $10^4$ are Lys57, Glu63, Arg83, and Leu86 in the A* monomer and Val64, Lys66, Tyr85, Leu86, Asn87, Met88, and Met108 in the B* monomer. }
  \label{fig:MI2}
  \end{center}
\end{figure}

TR binding changes the clustering, particularly of the FG loops.  In CC:TR both the A* and B* FG loops are clustered separately from the main body of the protein and regions in direct contact with TR.  Trp82, which was found by the string calculations to be crucial for the conformational transition, is also clustered separately from the main protein body in both A* and B* conformations.  This implies that the B* FG loop is relatively independent from the TR-bound residues in CC:TR.  However, in AB:TR the B* FG loop is clustered with the main body of the protein that is in contact with TR.  Hence, the conformation of the B* FG loop in AB:TR is modulated by the bound TR.  Recall that the RMSF results showed that the B* FG loop in AB:TR is much more confined than in the AB system.  These two results together strongly suggest that the B* FG loop's dynamics is modulated by the TR in the AB:TR conformation.   The AB:TR system also stands out as it has higher total MI in the system (Fig.~\ref{fig:MI2}a) compared to the other three dimers.  Hence, not only does the clustering and RMSF data show that TR binding plays a key role in modulating the B* FG loop through residue correlations in AB:TR, but also that the overall intra-protein communication is stronger in AB:TR than in the other three systems.

% FIGURE

{\bf Betweenness centrality identifies communication pathways.} To gain a molecular-scale understanding of how residue conformational information is transferred across the protein, we filtered the complete mutual informational data to include only pairs of residues in direct contact, defined as residues that are 5.5\AA\  or closer for 75\% of the simulation time.   We then used this contact-filtered MI graph to calculate the betweenness centrality for each residue as discussed in Sec.~\ref{subsec:mut_info}.  Betweenness centrality represents the number of shortest paths between all node pairs on the MI graph that pass through a given residue, and is thus a measure of how important each residue is for information flow through the network.  The betweenness centrality has been used in previous work to determine the importance of nodes in networks constructed from residue-residue interaction energies \cite{Ortiz2015}.   Here, the network is constructed using residue-residue correlations.  In Fig.~\ref{fig:MI2}b we show the 11 residues with the highest betweenness centrality ($>10^4$) for the AB:TR system. We find that there is a group of consecutive residues with high centrality on the B* G $\beta$-strand (Tyr85, Leu86, Asn87 and Met88), including Tyr85 which has highest centrality in the system (Fig.~\ref{fig:MI2}c).  Two other residues with high centrality are Val64 on the B* F $\beta$-strand, with the second highest centrality, and Lys66 in the B* FG loop.  These results thus suggest a communication network in the protein in which the majority of communication travels through the spine of four residues on the B* G $\beta$-strand and is then coupled to the B* F $\beta$-strand and FG loop. Perturbations due to TR binding passed along this pathway are thus strongly coupled to the FG loop conformation. In contrast, the same group of residues does not exhibit high centrality in the A* monomer, providing an explanation for why TR binding does not affect the A* FG loop conformation.

\begin{figure}[hbtp]
  \begin{center}
    \includegraphics[width=\columnwidth]{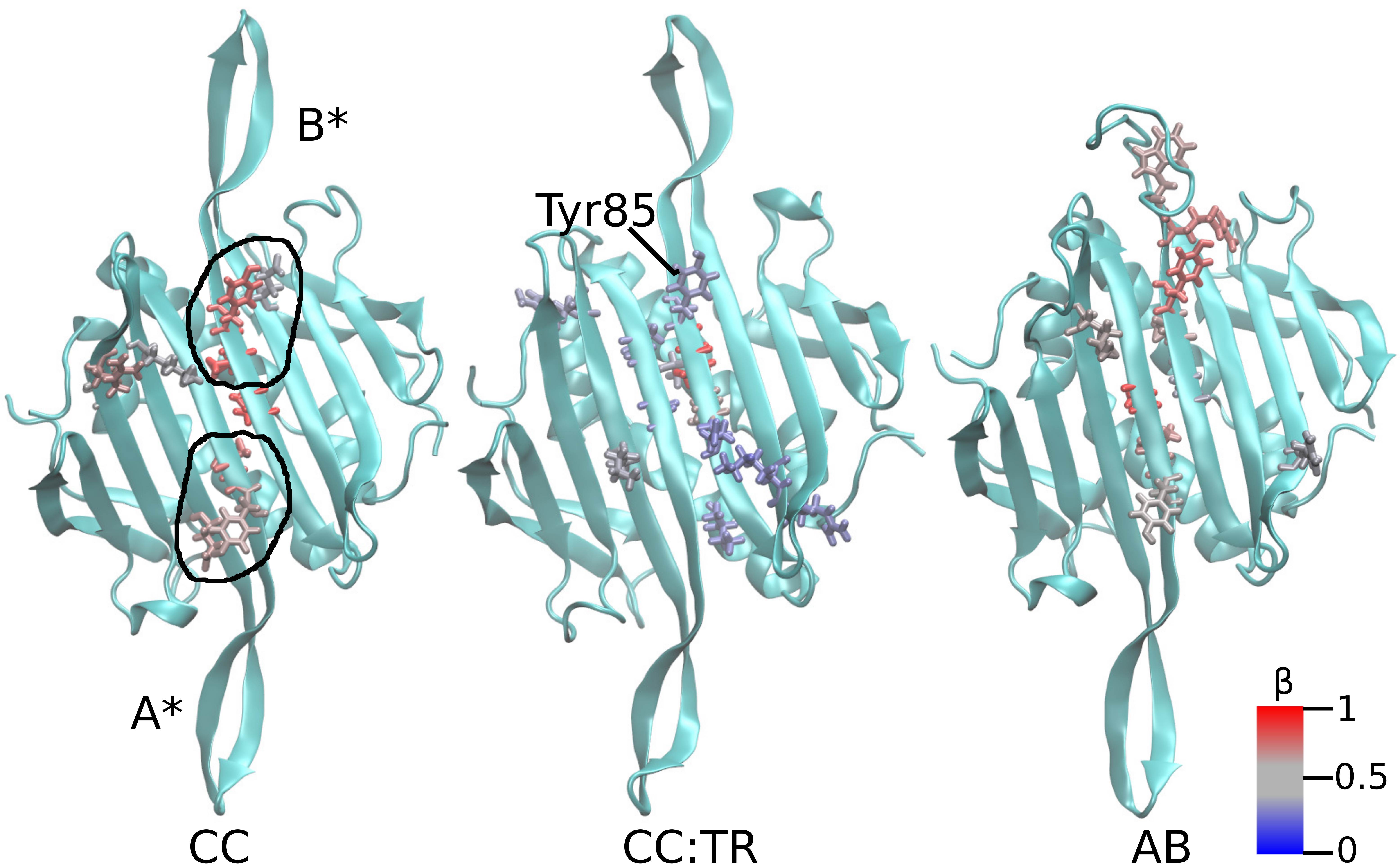}
   \caption{The relative betweenness $\beta$, as described in Fig.~\ref{fig:MI2}, shown for residues with a betweenness centrality greater than $10^4$ for the CC, CC:TR and AB systems.  The three residues Glu63, Tyr85 and Leu86 that appear on both the A* and B* monomers of the symmetric CC dimer are outlined.  Tyr85 which appears as a high betweenness residue in the B* monomer of all of the $\cp$ and $\cp$:TR systems is marked on the CC:TR structure.}
  \label{fig:MI3}
  \end{center}
\end{figure}

Considering the residues with high centrality ($>10^4$) in the other three dimer systems gives further insight into how intra-protein communication is modulated by TR binding (Fig.~\ref{fig:MI3}).  For the CC dimer, there is a group of 3 residues (Glu63, Tyr85, and Leu86), that appear in both the A* and B* monomers and are outlined in Fig.~\ref{fig:MI3}.  These residues highlight the symmetry that exists in the CC dimer.  However, the highest centrality residues are not completely symmetric between the A* and B* monomer.  For example, residue Met88 appears in the B* monomer but not in the A* monomer.  We expect these asymmetries to be resolved with greater sampling, but this result also highlights the fact that the instantaneous configurations of the CC dimer are not perfectly symmetric due to fluctuations.  The CC:TR results show that TR binding breaks the symmetry between the A* and B* monomers present in the CC dimer, since the high centrality residues present in the B* monomer are not present in the A* monomer.  However, there is a similarity between the B* monomers in CC and CC:TR, as residues Tyr85, Leu86 and Met88 in the G  $\beta$-strand appear in both dimers.  Tyr85, Leu86 and Met88 also appear in both the A* and B* monomers of the AB dimer and the B* monomer of AB:TR.  Hence, these three residues are important for communication in the B* monomer in all of the dimer systems.

% FIGURE

\section{Discussion and Conclusions}
\label{sec:discussion}

We have combined the string method, free energy calculations, and analysis of long unbiased molecular dynamics simulations to characterize the effect of binding of the MS2 genome fragment TR to its capsid protein.  The calculations demonstrate that the impact of TR binding is substantial and far-reaching. The free energy profiles calculated from our converged strings for the \ABS and \ABTR transitions (Fig.~\ref{fig:ms2_free_energy}) show a strong shift in the favored population from CC to AB:TR.  Furthermore, the strings indicate that TR binding dramatically alters the interconversion pathway, changing the sequence of events and the nature and number of intermediate metastable states.  Given that TR binds more than a nanometer from the the residues which undergo the majority of conformational rearrangement (the FG loop), our calculations provide direct evidence for allostery and begin to reveal its underlying mechanisms, albeit within the limitations of force field accuracy and finite sampling.

The fundamental effect of TR-binding is to generate an inherently asymmetric dimer.  The CC$\rightarrow$AB transition requires a spontaneous fluctuation that breaks the CC symmetry and `chooses' which FG loop will interconvert to a B conformation. In contrast, TR-binding introduces subunit-spanning asymmetries that favor transition of one chain. We characterized these asymmetries, and how they are transmitted across the protein, by analyzing collective motions and correlated conformational statistics of amino acids within long unbiased MD trajectories of each stable substate. We found extensive asymmetries in both the dynamical fluctuations and correlations.  The most significant effect of TR binding was found to be on the FG loop of the B* monomer when comparing the AB and AB:TR conformations.  We find a pathway of strong communication along a spine of residues between the TR binding region and the B* FG loop, thus identifying how the conformational landscape can be so strongly shifted upon TR binding to stabilize the AB conformation.

\textbf{Comparison to previous results.}
Previous experiments on MS2 have shown that TR binding induces a conformation change from a symmetric to an asymmetric structure~\cite{Stockley2007}.  Based on this and other evidence it has been inferred that the CC state is preferred in the absence of TR, and the AB state in the presence of TR.  Our results from the string method calculation and the associated free energy profile directly support this conclusion, and also reveal that the associated transition pathways differ in the presence of TR.

A study by Dykeman et al\cite{Dykeman2010} performed an all-atom normal mode analysis to determine how the vibrational modes are modified by RNA binding.  They found that TR binding to a (WT) CC conformation causes asymmetric fluctuations of the EF loop; fluctuations increase in A$^*$ and decrease in B$^*$.  The mutant Trp82Arg has an asymmetry in the DE loop instead, which was proposed as a possible explanation for why it is assembly-incompetent.
Our MSF calculations also find asymmetries upon TR binding in the EF loop.  While there is little difference between the FG loop fluctuations in the CC and CC:TR systems, TR binding in the AB systems shows the fluctuations in the B* FG loop greatly decreased upon TR binding.

\textbf{Limitations of our calculations and outlook.}
The relevance of the string method pathway and associated free energy profile depend on the extent to which the collective variables describe all relevant slow degrees of freedom.  Furthermore, recent computational studies have shown that conformational transitions can proceed by multiple, diverse pathways (\eg \cite{Pontiggia2015}), while a single string calculation typically samples only one transition tube.
We have assessed several metrics to determine whether our set of collective variables was sufficient and the extent of sampling within trajectory space:   (1) Convergence during string iterations was reasonably rapid.  Missing slow degrees of freedom can be expected to slow convergence since they relax on a slow time scale. (2) The umbrella sampling calculations were performed in both directions along the pathway.  A lack of significant hysteresis between these two directions is consistent with inclusion of all relevant degrees of freedom. (3) Independent string calculations started from substantially different initial pathways led to very similar converged strings (Fig.~\ref{fig:ms2_string_validation2}), consistent with efficient sampling in trajectory space. Taken together, these results are consistent with a sufficient set of collective variables and broad sampling within trajectory space. Further testing of the validity of these calculations could be achieved by comparing the results to those of an independent technique, such as Markov State model (MSM) calculations.  It is also possible to use the converged string as a starting point for efficient construction of an MSM \cite{Pontiggia2015}.

While we analyzed correlations of amino acid conformations within the stable conformational substates, further insight into the transition mechanism might be obtained by characterizing mutual information during the conformational transition.  Finally, in this work, we focused on the effects of TR binding on the coat protein dimer conformations. A natural next step is to examine the effect of TR binding on dimer-dimer interactions; \eg Ref. \cite{ElSawy2010}.

\begin{acknowledgments}
This work was supported by the NIH through Award Number R01GM108021 from the National Institute Of General Medical Sciences (MRP and MFH)and R01GM100966 from NIGMS (DTM and MFH).  Computational resources were provided by the NSF through XSEDE computing resources (Trestles, Kraken, Queen Bee, and Maverick) and the Brandeis HPCC which is partially supported by the Brandeis Center for Bioinspired Soft Materials, an NSF MRSEC, DMR-1420382.
\end{acknowledgments}

%%% Appendix %%%
\appendix

\setcounter{figure}{0}
\renewcommand{\thefigure}{A\arabic{figure}}

\section{Additional methods details}
\label{appendix:ms2_simulations}

\subsection{Equilibration}
\label{appendix:subsec:ms2_equilibration}
Each of the 4 systems was relaxed from its initial configuration as follows.  First, the system was minimized while iteratively relaxing harmonic restraints on all protein heavy atoms, centered on crystal structure positions. Next, MD simulations were performed in which the same restraints, now centered on the final minimized position, were slowly relaxed as the temperature was gradually increased from 25K to 300K.  Unbiased MD was then performed for 50-100ns to ensure equilibration.  To prevent self-interaction, rotational drift was limited by harmonic restraints on the $\alpha$-carbons of residues 105-109 in the A subunit $\alpha$-helix, which is located on the top of the dimer far from the FG-loop and RNA binding sites.

%(Fig.~\ref{fig:ms2_equil_ab})
Of the four systems, only the AB dimer undergoes significant rearrangement.  Trp82 in the FG loop rotated and the rest of the FG loop moved toward the EF hairpin of the A monomer; in contrast, the AB:TR FG loop remained close to the DE loop of the B monomer.  The difference between equilibrated AB and AB:TR states is significant because Trp82 is a large side chain which rearranges substantially during the conformational changes.

\section{String method calculations}
\label{appendix:string}
This section presents details on the string method calculation.
Following Ovchinnikov et al. \cite{Ovchinnikov2011}, we define $N_{\mathrm{cv}}$ collective variables that depend on the Cartesian positions $\mathbf{x}$ of atoms in the protein as  $\hat{\boldsymbol{\theta}}(\bf{x}) = \left(\hat{\theta}_1(\bf{x}), \hat{\theta}_2(\bf{x}), \dots, \hat{\theta}_{N_{\mathrm{cv}}}(\bf{x})\right)$.  Each image $n$ of the string evolves according to\cite{Ovchinnikov2011}
\begin{equation}
\label{eq:string_evolve}
\boldsymbol{\theta}_n(t+\Delta t) = \boldsymbol{\theta}_n(t) - \gamma^{-1} \Delta t \mathbf{M}(\boldsymbol{\theta}_n(t)) \nabla G(\boldsymbol{\theta}_n(t))
\end{equation}
where $\boldsymbol{\theta}_n(t)$ gives the collective variable values of image $n$ from string iteration $t$ and $\gamma$ is a tuneable ``friction constant'' that sets the size of the step taken down the free energy gradient (along with $\Delta t$).  The metric tensor $\mathbf{M}(\boldsymbol{\theta}(t))$ accounts for the curvilinear nature of the collective variables and is given by\cite{Ovchinnikov2011}
\begin{equation}
\label{eq:string_metric_tensor}
M_{ij}(\boldsymbol{\theta}) = \sum_k \frac{1}{m_k} \Bigg \langle  \frac{\partial \hat{\theta_i}(\mathbf{x})}{\partial x_k} \frac{\partial \hat{\theta_j}(\mathbf{x})}{\partial x_k} \Bigg \rangle_{\hat{\boldsymbol{\theta}}(x)=\boldsymbol{\theta}}
\end{equation}
where the sum ranges over each coordinate $k$ for all atoms in the system, $\langle \mathrm{\dots} \rangle$ denotes an average over sampling constrained in the vicinity of $\boldsymbol{\theta}$, and  $m_k$ is the mass of atom $k$.

We made two simplifying approximations in our implementation.  Since we used only Cartesian coordinates for collective variables, we approximated the tensor $M(\boldsymbol{\theta}_n(t))$ in Eq.~\eqref{eq:string_evolve} as the identity matrix.  Tests with and without this approximation supported that the metric tensor can be neglected for our system.

 The second approximation was to dynamically set $\gamma^{-1} \Delta t$ (from Eq.~\eqref{eq:string_evolve}) such that the step size is a fixed fraction of the image spacing.   This guarantees that new images will not jump too far in any given iteration.  With the alanine dipeptide model system, we extensively tested our implementation with both approximations against a string implementation with collective variables based on dihedral angles.

To identify collective variables sufficient to describe the transition between states, we  systematically vetted candidate coordinates using restrained targeted molecular dynamics simulations~\cite{Ovchinnikov2011} (described for our systems in Appendix~\ref{subsubsec:selecting_CVs}).  Next, we used TMD to generate an initial string connecting the two metastable states.  This pathway was then discretized into images, and the string was systematically relaxed by the following iterative procedure.

\begin{enumerate}
\item \textbf{Sample}.  For each image $n$, run short simulations to estimate $\nabla G(\boldsymbol{\theta}_n)$ (the free energy gradient in collective variable space in the proximity of image $n$).  In each short simulation,  impose a harmonic potential for each collective variable,  centered on the image.  The spring constant of the harmonic potential is selected to keep sampling in the vicinity of its image (typically with an average sampling radius of 1-2 image spacings).  Calculate the average force imposed by each potential.

\item \textbf{Evolve}.  Generate a new string by displacing each image a distance  $\delta$ in the direction opposite to the free energy gradient. In our implementation, $\delta$ is scaled to be a fixed fraction of the image spacing.

\item \textbf{Reparameterize}.  Redefine the locations of images along the string so that they are uniformly spaced in arc length, following the implementation of Maragliano et al. \cite{Maragliano2006}.
\end{enumerate}

This procedure was iterated until the string pathway approximately converged, which was assessed by the RMSD between the initial and current strings.  We define the RMSD between strings as
\begin{equation}
\label{eq:string_rmsd}
\mathrm{RMSD}(\boldsymbol{\theta}^{\str_1},\boldsymbol{\theta}^{\str_2}) = \left( \int_0^1 \! |\boldsymbol{\theta}^{\str_1}(\alpha)-\boldsymbol{\theta}^{\str_2}(\alpha)|^2 \mathrm{d}\alpha \right)^{1/2}
\end{equation}
with $\boldsymbol{\theta}^{\str_i}(\alpha)$ as the $N_\mathrm{cv}$-dimensional point at fraction $\alpha$ along string $\str_i$.  Since the strings are discretized, it is necessary to interpolate between images.

\subsection{Generating the Initial String}

Initial strings were generated from TMD trajectories of the CC$\rightarrow$AB and CC:TR$\rightarrow$AB:TR transitions, with the TMD bias based only on CV atoms. Coordinates were saved every 2ps and used to construct a time series of CV values, which was then smoothed to prevent noise from dominating image selection.  The data was smoothed by applying a nearest neighbor smoothing kernel to the coordinates for 10-20 iterations.    Forty images, with approximately equal spacing, were then selected from the smoothed trajectory and used as the initial string pathway.  The spacing between the $\Ncv$-dimensional images in the initial string was $3.5$\AA\ for $\cp$ and $2.5$\AA\ for $\cp$:TR, which provided sufficient resolution to capture bond-breaking and all significant conformational rearrangements. %

\textbf{TMD parameters.}  During selection of collective variables and generation of initial pathways, TMD simulations imposed a harmonic potential as a function of the RMSD difference between the current and target structure, measured from the positions of the candidate CV atoms only. The center of the potential was moved linearly from the RMSD of the initial configuration to 0 over 1.5 ns. The spring constant was linearly scaled from $k = 2.5\times 10^6 \mathrm{kJ}/\mathrm{mol} \cdot \mathrm{nm}^2$ to $k = 5\times 10^6 \mathrm{kJ}/\mathrm{mol} \cdot \mathrm{nm}^2$ over this same interval.  After centering on RMSD$= 0$, $k$ was linearly increased over three separate 500ps intervals to $k = (2,20,200)\times 10^7 \mathrm{kJ}/\mathrm{mol} \cdot \mathrm{nm}^2$.  
After this, $k$ was linearly decreased to 0 over 1ns, followed by 4ns of unbiased simulation.

\subsection{Running the String}
\label{appendix:subsec:running_string}

Each string was evolved according to the three steps outlined at the beginning of Appendix~\ref{appendix:string}: sample, evolve, reparameterize.  In the sample step, for each image $n$, the structure from the previous (or initial) string closest in CV space to the image CV values $\boldsymbol{\theta}_n$ was subjected to a steered molecular dynamics (SMD) simulation targeting  $\boldsymbol{\theta}_n$.  A harmonic potential with force constant $k_{\mathrm{drag}} = 10^5$kJ/nm$^2$ was imposed for each CV, and moved linearly to  $\boldsymbol{\theta}_n$ (at a speed of no faster than $0.1$nm/(1000steps)).  Then, we sampled the local free energy gradient by performing MD for an additional $200$ps with harmonic restraints for the CV centered at $\boldsymbol{\theta}_n$. To maintain local sampling while speeding convergence, we chose a restraint force constant of $k_{\mathrm{hold}} = 450.0$kJ/nm$^2$, yielding an average sampling radius of 1-2 image spacings. CV values were recorded every $0.1$ps. The string was then evolved by updating CV values according to Eq.~\ref{eq:string_evolve}, with a step size set to a fixed fraction $\delta=0.5$ of the image spacing, followed by reparameterization to maintain uniform spacing along the arc length.

% FIGURE
\begin{figure}[tbp]
  \begin{center}
    \includegraphics[width=\columnwidth]{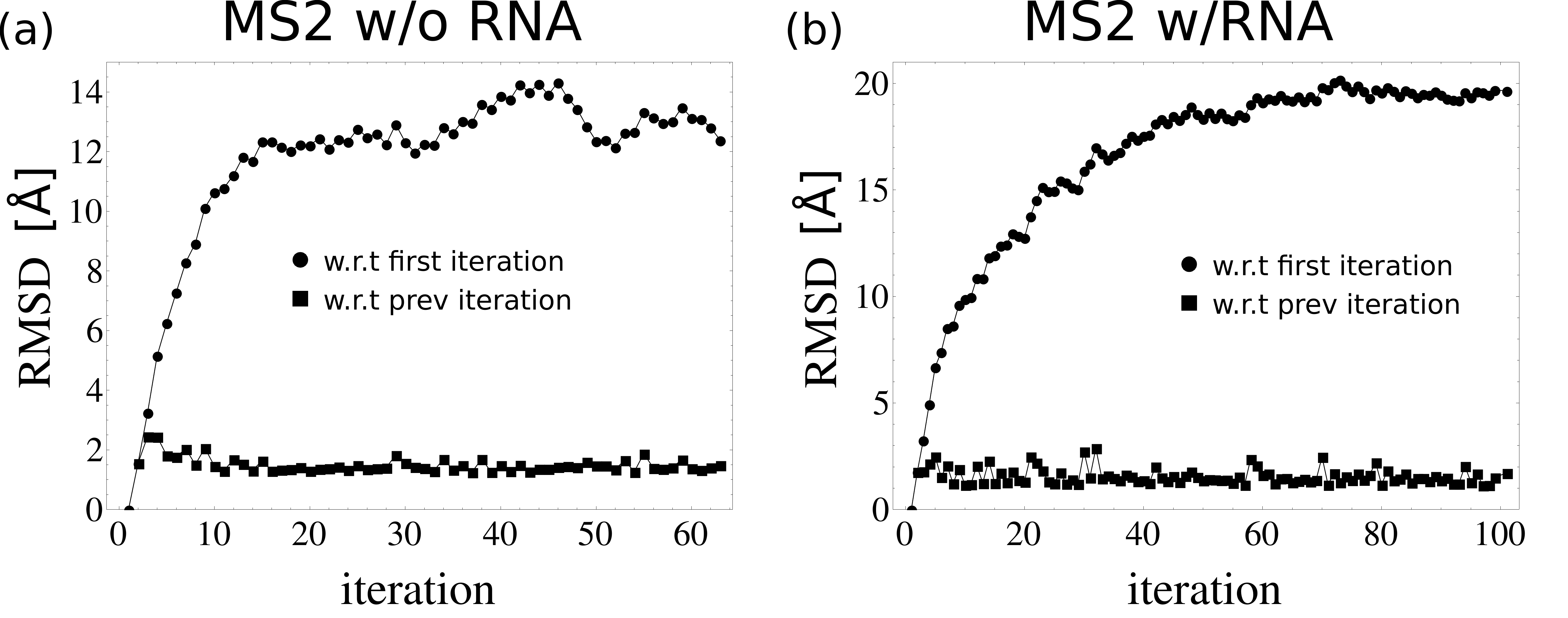}
  \caption{The RMSD during the equilibration of the {\bf (a)} $\cp$ and {\bf (b)} $\cp$:TR strings.  The RMSD is calculated according to Eq.~\eqref{eq:string_rmsd} with respect to both the initial string and the previous string iteration.}
  \label{fig:ms2_string_rmsd}
  \end{center}
\end{figure}

To monitor string convergence, we calculated the RMSD (in CV space) between points at equal arc length along the string according to Eq.~\ref{eq:string_rmsd}.  We used a linear interpolation between neighboring images to calculate the RMSD at arc lengths not commensurate with image locations.  Each string was run until the RMSD with respect to the initial string plateaued, which required 50-100 iterations (Fig.~\ref{fig:ms2_string_rmsd}).

\subsection{String Convergence and Validity}
\label{appendix:string_convergence}
\begin{figure}[btp]
  \begin{center}
    \includegraphics[width=\columnwidth]{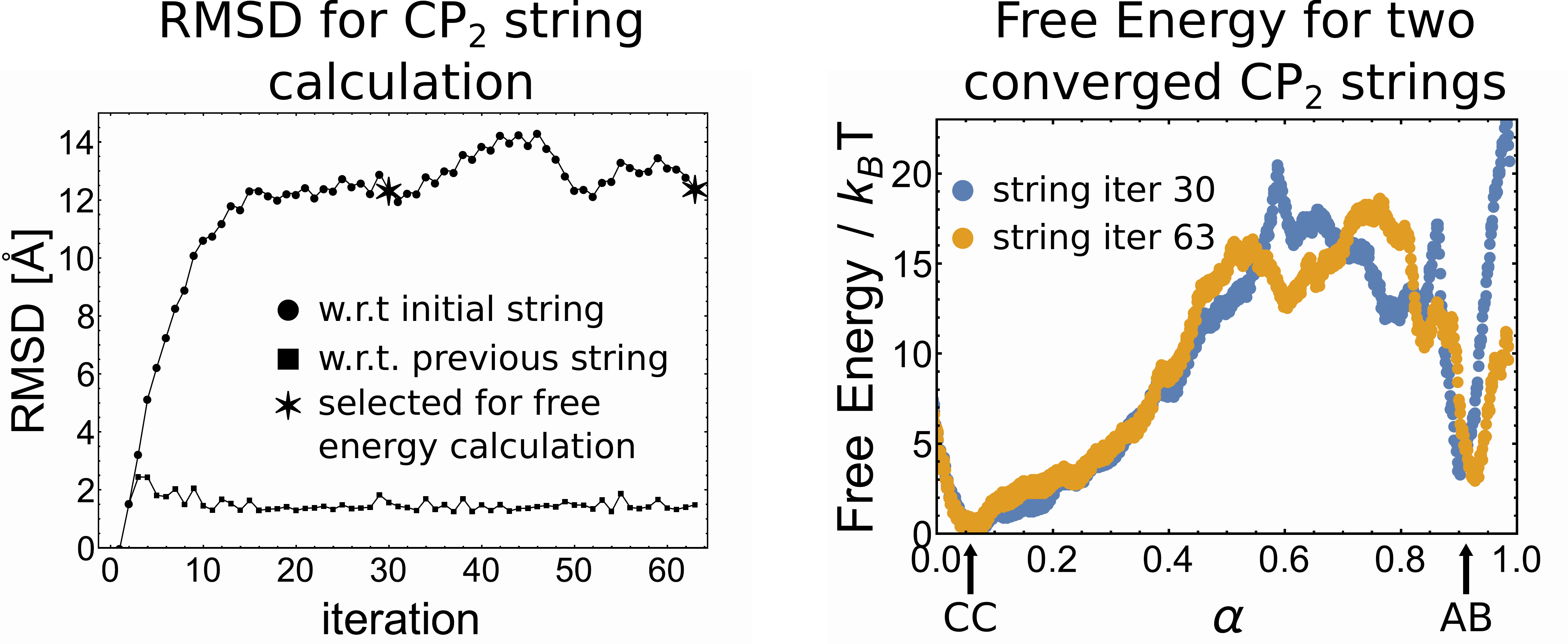}
  \caption{\textbf{(a)} The string RMSD as calculated from Eq.~\eqref{eq:string_rmsd} during the convergence of a $\cp$ string.  The strings at iterations 30 and 63 were taken for the free energy calculation (as marked by the stars).  \textbf{(b)} The free energy profile for each of the two converged $\cp$ strings as a function of arc length $\alpha$.}
  \label{fig:ms2_string_validation1}
  \end{center}
\end{figure}

To test the assumption that a plateau of the RMSD in CV space is a good measure of string convergence, we calculated the CC$\rightleftharpoons$AB free energy profile for two string iterations after the RMSD plateau (Fig.~\ref{fig:ms2_string_validation1}). Although the two free energy profiles are not identical, they obtain the same free energy difference between CC and AB substates, nearly the same barrier height, and both have a single on-pathway metastable state. 

% FIGURE

\begin{figure}[btp]
  \begin{center}
    \includegraphics[width=\columnwidth]{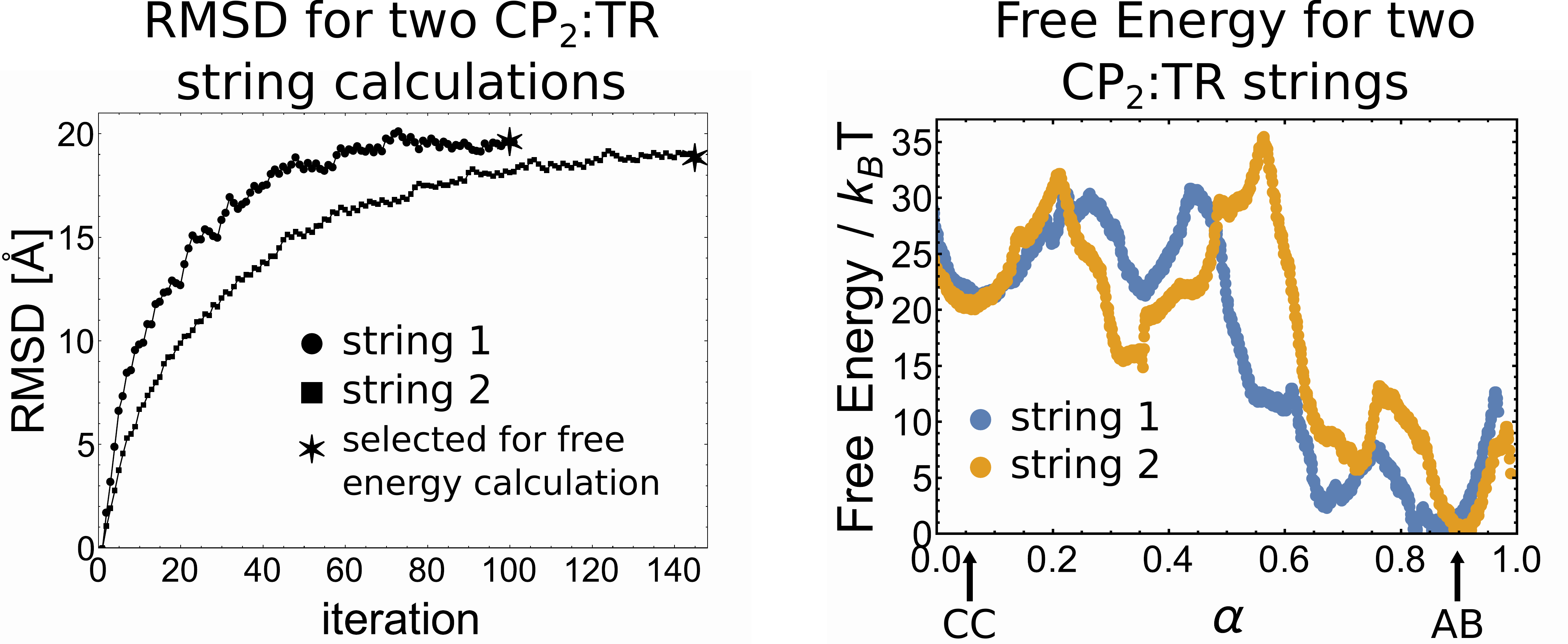}
 \caption{\textbf{(a)} The RMSD with respect to the initial string (Eq.~\ref{eq:string_rmsd}) during the convergence of $\cp$:TR strings initialized from two independent TMDs.  \textbf{(b)} The free energy profiles for the two final $\cp$:TR strings as a function of arc length $\alpha$.} \label{fig:ms2_string_validation2}
  \end{center}
\end{figure}

To assess global convergence of the string, we performed a second string calculation for the CC:TR$\rightleftharpoons$AB:TR transition, initialized from a different TMD simulation. This TMD used a slightly different definition for the CC:TR and AB:TR substates, and produced an initial pathway which differs substantially from the initial pathway used for the first string. The RMSD of $13\AA$ between the two initial pathways is approximately as large as the RMSD between the first converged string and its initial pathway. The convergence and resulting free energy profiles are shown in Fig.~\ref{fig:ms2_string_validation2}.  Once again, the two strings result in the same relative free energies for the CC:TR and AB:TR substates and contain the same number (two) of on-pathway metastable states. While there are quantitative differences, the overall similarity between the two calculations suggests that the strings have converged to the same pathway.
This result from two different initial pathways is consistent with a global MFTP, although a thorough assessment would require a number of additional strings and hence a large computational cost.

% FIGURE

\section{RMSF}
\label{appendix:rmsf_results}

\begin{figure}[btp]
  \begin{center}
    \includegraphics[width=\columnwidth]{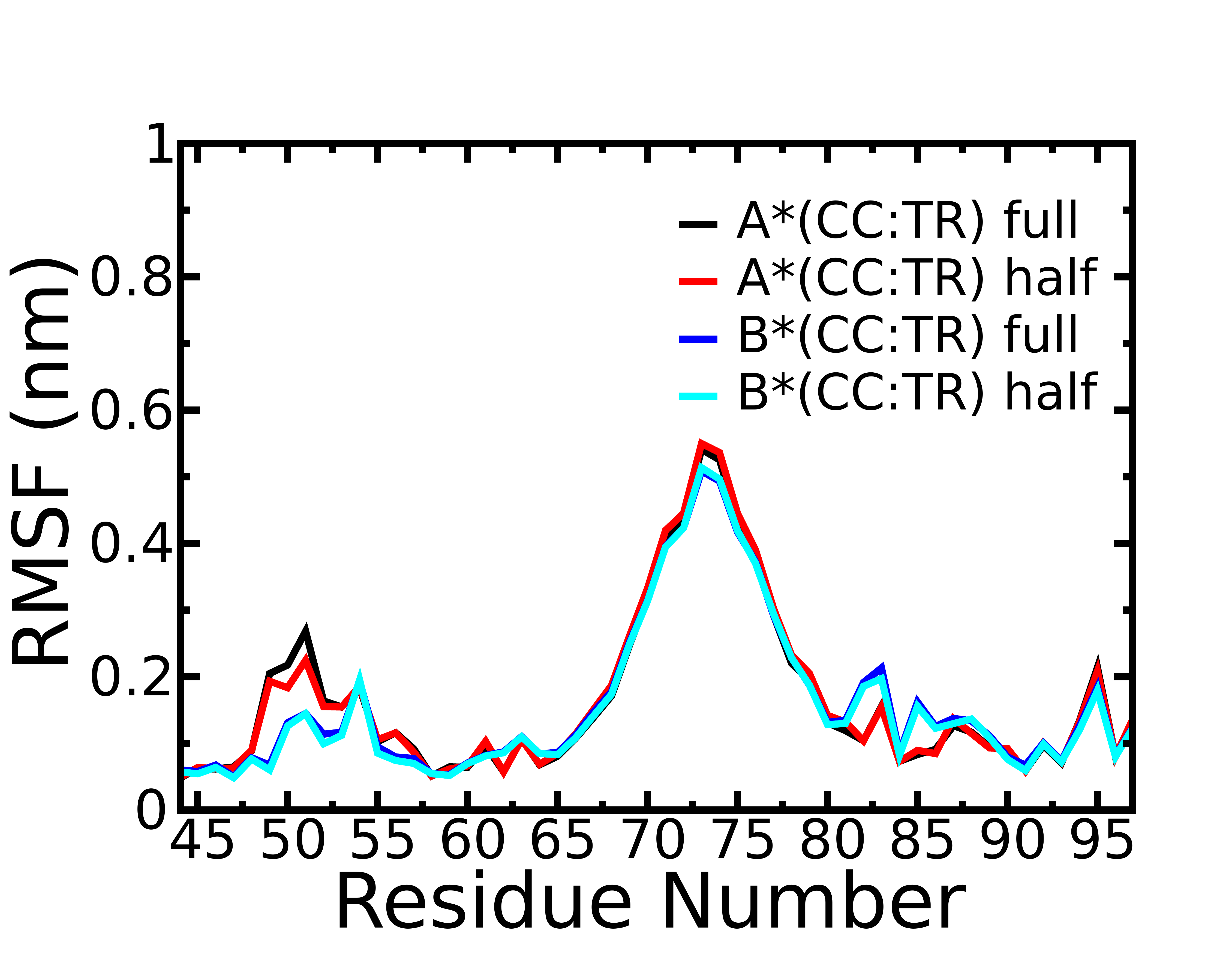}
  \caption{The per residue RMSFs of the EF and FG loop regions of the A* and B* monomers of the CC:TR dimer.  The results using the full simulation data and half the simulation data are compared and show that the results obtained using the full simulation data are converged.}
  \label{fig:rmsf_CC-TR_Full-half}
  \end{center}
\end{figure}

To validate that the sampling used is sufficient for the RMSF calculations we compare the per residue RMSF for the EF and FG loop regions for the CC:TR system using the full data set of 2.576$\mu$s and using half the total trajectory frames.  These results are presented in Fig.~\ref{fig:rmsf_CC-TR_Full-half} and are marked ``full'' and ``half''.  The RMSF values obtained from using the full data set are the same as what was presented in Fig.~\ref{fig:msf}a.  The results show that the two data sets give very similar results and that the calculated RMSF values are converged.
% FIGURE

\section{Mutual Information Calculations}
\label{appendix:mutinfo_results}

To determine groups of amino acids that have correlated distributions, we used hierarchical clustering on the MI matrices.  In hierarchical clustering, each amino acid starts in its own cluster, and clusters with minimal ``dissimilarity'' are recursively merged until only one cluster remains.  For our calculation, the ``dissimilarity'' was determined by the intra-cluster average of $D_{ij}$ of Eq.~\eqref{eq:dissim_mut_inf}.  From the resulting hierarchy of clusters, we systematically extracted the largest possible clusters, such that the intra-cluster MI average was greater than a certain cutoff value.  Hierarchical clustering was applied to each of the four MI data sets.  An intra-cluster average cutoff of 0.035$\kt$ was used for the CC system, which results in 7 distinct clusters.  Increasing the cutoff to 0.04$\kt$ lead to a jump in the number of clusters to 12, hence we decided to base our analysis on the smaller 7-cluster result.  A similar approach was used for the other three dimer systems, where a cutoff was found to generate a reasonable number of clusters with low inter-cluster MI averages, but stayed below an intra-cluster average cutoff that would lead to a jump in the number of clusters.  Hence, an intra-cluster average cutoff of 0.04$\kt$ was used for the AB system, which results in 5 distinct clusters.  Increasing the cutoff to 0.05$\kt$ lead to an increase in the number of clusters to 10.  For the CC:TR system an intra-cluster average cutoff of 0.07$\kt$ was used, resulting in 8 distinct clusters.  Increasing the cutoff to 0.08$\kt$ lead to an increase in the number of clusters to 14.  For the AB:TR system an intra-cluster average cutoff of 0.10$\kt$ was used, resulting in 5 distinct clusters.  Increasing the cutoff to 0.11$\kt$ lead to an increase in the number of clusters to 10.

\begin{figure}[btp]
  \begin{center}
    \includegraphics[width=\columnwidth]{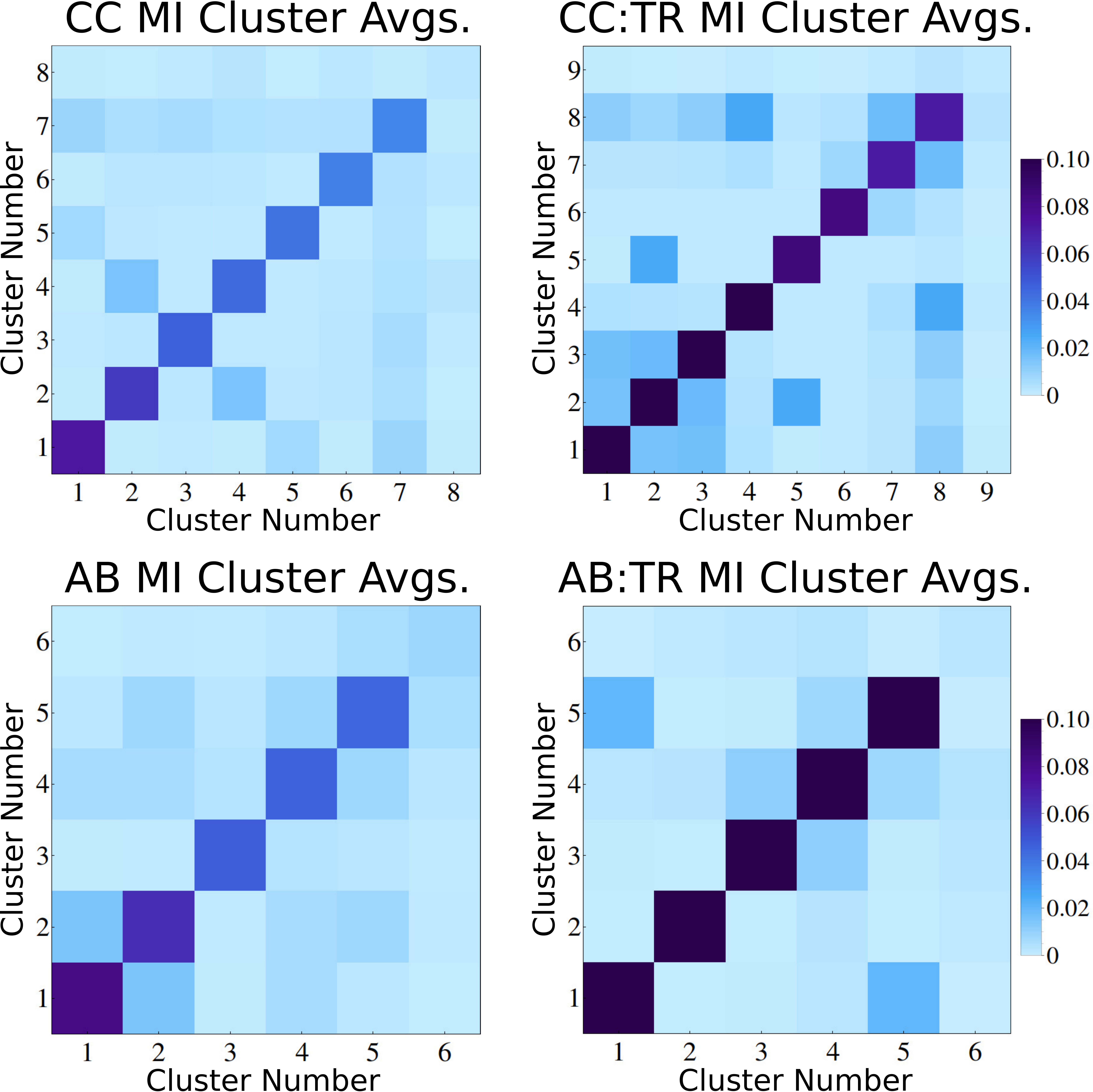}
  \caption{The similarity matrix for the mutual information matrix clusters for each $\cp$ and $\cp$:TR system.  The coloring indicates the average mutual information and is scaled uniformly for all plots (as shown on the right in units of $\kt$).  The last cluster in each plot contains amino acids that do not share MI with each other or other clusters, which is why the last diagonal element has a value near 0.}
  \label{fig:mutinf_valid}
  \end{center}
\end{figure}

The resulting clusters presented in Fig.~\ref{fig:MI1} were tested to ensure that they have a high intra-cluster correlation average and a low inter-cluster correlation average (as calculated from Eq.~\ref{eq:clusterAvgs}).  The resulting correlations are shown in Fig.~\ref{fig:mutinf_valid}, where the diagonal shows strong intra-cluster correlations.

% FIGURE

\begin{figure}[btp]
  \begin{center}
    \includegraphics[width=\columnwidth]{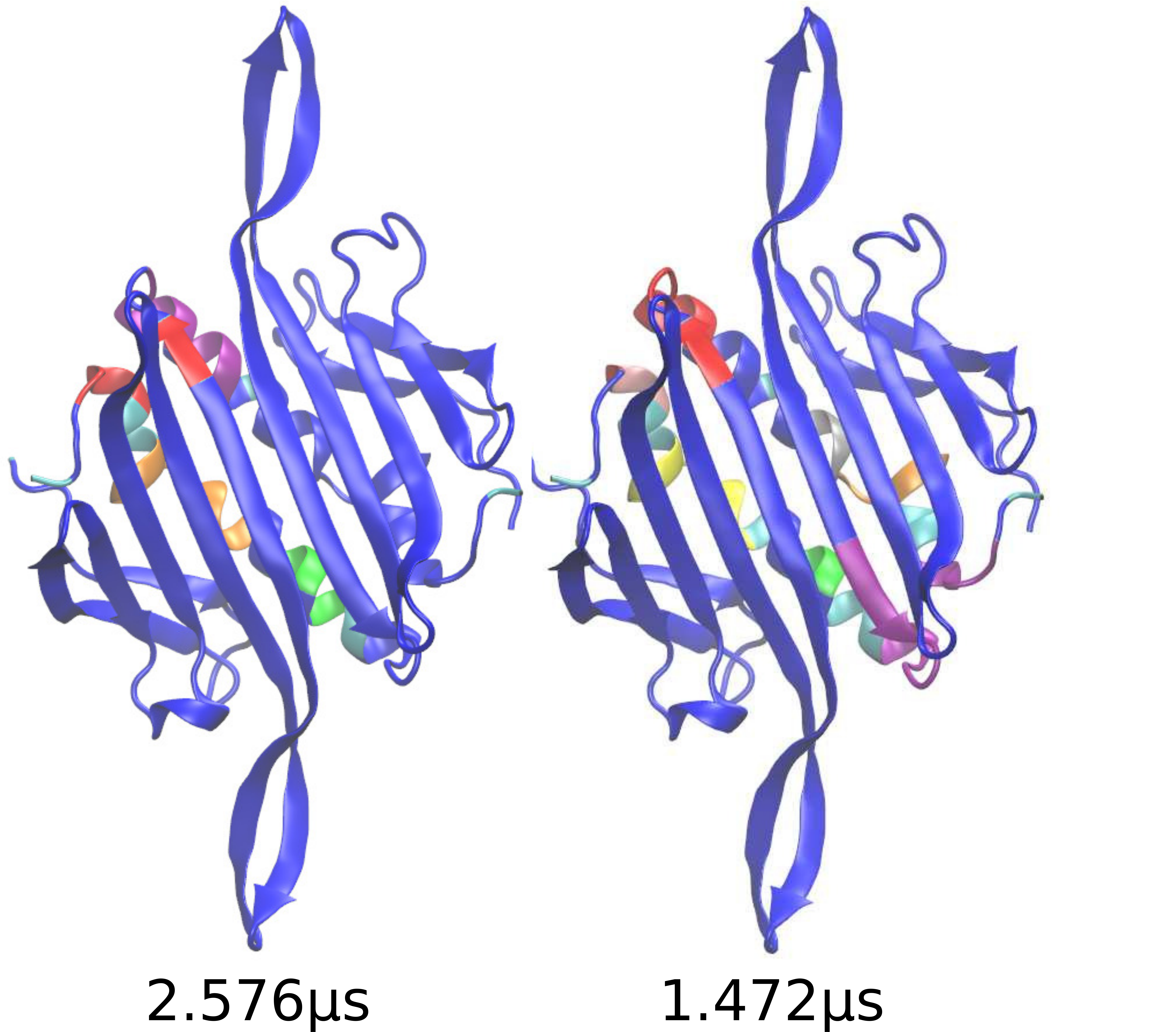}
  \caption{The clusters of the MI matrix for the CC system using the 2.576$\mu$s and 1.472$\mu$s of sampling.  The similarity in the resulting clusters indicates that the sampling is converged.}
  \label{fig:mutinf_valid2}
  \end{center}
\end{figure}

The MI matrix for the CC dimer was clustered using all of the 7 trajectories (2.576$\mu$s) as shown in Fig.~\ref{fig:MI1} and using 4 trajectories (1.472$\mu$s).  The resulting clusters for both sampled MI data sets for CC are compared in Fig.~\ref{fig:mutinf_valid2}.  Using an intra-cluster average cutoff of 0.05$\kt$  lead to 8 distinct clusters.  The cluster identifications are similar in for the 2.576$\mu$s and 1.472$\mu$s sampled systems.  Importantly, the FG loops in the A* and B* monomers are clustered with the large blue cluster that includes the TR binding sites.  Hence, we conclude that the sampling is converged for the calculation of the MI clusters.

% FIGURE

\end{document}